\documentclass[a4paper,11pt]{article}
\pdfoutput=1 

\usepackage{jinstpub} 

\usepackage{lineno}
\title{\boldmath Characterization of BNL and HPK AC-LGAD sensors with a 120 GeV proton beam}

\usepackage{siunitx} 
\DeclareSIUnit\sq{\ensuremath{\Box}}                           

\usepackage[rawfloats=true]{floatrow}

\usepackage{subcaption}
\makeatletter
\renewcommand\p@subfigure{\thefigure.}                        
\makeatother

\usepackage{xcolor}

\author[a]{Ryan Heller,}
\author[a,1]{Christopher Madrid%
\note{Corresponding author.},}\emailAdd{cmadrid@fnal.gov}
\author[a]{Artur Apresyan,}
\author[d,g,h]{William K. Brooks,}
\author[b]{Wei Chen,}
\author[b]{Gabriele D'Amen,}
\author[b]{Gabriele Giacomini,}
\author[f]{Ikumi Goya,}
\author[f]{Kazuhiko Hara,}
\author[f]{Sayuka Kita,}
\author[a]{Sergey Los,}
\author[a,c]{Adam Molnar,}
\author[e]{Koji Nakamura,}
\author[a]{Cristián Peña,}
\author[d,g]{Claudio San Martín, }
\author[b]{Alessandro Tricoli,}
\author[f]{Tatsuki Ueda,}
\author[a,c]{Si Xie}

\affiliation[a]{Fermi National Accelerator Laboratory, \\PO Box 500, Batavia IL 60510-5011, USA}
\affiliation[b]{Brookhaven National Laboratory,\\Upton, 11973, NY, USA}
\affiliation[c]{California Institute of Technology,\\Pasadena, CA, USA}
\affiliation[d]{Departamento de F\'isica y Astronom\'ia, Universidad Técnica Federico Santa María, \\Valparaiso, Chile}
\affiliation[e]{High Energy Research Organization,\\ Oho 1-1, Tsukuba, Ibaraki, 305-0801, Japan}
\affiliation[f]{University of Tsukuba,\\ 1-1-1 Tennodai, Tsukuba, Ibaraki, 305-8571, Japan}
\affiliation[g]{Centro Cient\'ifico Tecnol\'ogico de Valpara\'iso-CCTVal, \\
Universidad T\'ecnica Federico Santa Mar\'ia, Casilla 110-V, Valpara\'iso, Chile}
\affiliation[h]{Millennium Institute for Subatomic Physics at the High-Energy Frontier (SAPHIR) of ANID,\\Fern\'andez Concha 700, Santiago, Chile}

\abstract{
We present measurements of AC-LGADs performed at the Fermilab’s test beam facility using \SI{120}{\GeV} protons. 
We studied the performance of various strip and pad AC-LGAD sensors that were produced by BNL and HPK. 
The measurements are performed with our upgraded test beam setup that utilizes a high precision telescope tracker, and a simultaneous readout of up to 7 channels per sensor, which allows detailed studies of signal sharing characteristics. 
These measurements allow us to assess the differences in designs between different manufacturers, and optimize them based on experimental performance. 
We then study several reconstruction algorithms to optimize position and time resolutions that utilize the signal sharing properties of each sensor. We present a world's first demonstration of silicon sensors in a test beam that simultaneously achieve better than 6--10 \si{\um} position and \SI{30}{\ps} time resolution. This represents a substantial improvement to the spatial resolution than would be obtained with binary readout of sensors with similar pitch.
}


\begin{document}
\maketitle
\flushbottom

\section{Introduction}

Future proposed colliders require $\mathcal{O}(10)$~\si{\ps} time resolution and $\mathcal{O}(10)$~\si{\um} position resolution in order to disentangle the expected extreme number of simultaneous interactions per bunch crossing and improve particle identification capabilities~\cite{Sickling, Wulz277931111}. 
One of the major breakthroughs in detector technology in recent years has been the development of novel types of silicon detectors that provide good timing resolution. 
These detectors are based on technologies that have demonstrated timing resolution of few tens of picoseconds, for instance with Low Gain Avalanche Detectors (LGADs)~\cite{GIACOMINI201952, micronlgad, Pellegrini:2014lki, Paternoster_2017, GALLOWAY201919}. 
Developed for the High Luminosity LHC (HL-LHC) experiments, conventional LGADs provide precise time resolution of around \SI{30}{\ps}, but only coarse position resolution (approx. \SI{1}{\mm})~\cite{CMS:2667167, CERN-LHCC-2020-007}. 

The principle of operation of LGADs is based on the creation of a thin layer of high electric field at the $n^+$ and $p^+$ junction where the signal is amplified by avalanche multiplication. 
The LGADs for the HL-LHC are implemented as \SI{50}{\um} thick $n-$in$-p$ silicon sensors with a highly doped $p^+$ layer under the $n^+$ electrode. 
A prime challenge for these detectors is to minimize the inactive area between the pads to maximize the fill-factor, and consequently to achieve near $100\%$ detection efficiency. 
A significant reduction in LGAD cell sizes can be achieved by using the recently developed AC-coupled LGADs (AC-LGADs~\cite{ACLGADprocess, 8846722, RSD_NIM, firstAC}). These sensors achieve a $100\%$ fill-factor by using an uninterrupted gain layer and n+ implant, paired with finely segmented AC-coupled electrodes separated from the junction by a thin oxide layer. The charge is collected on the detector junction, and spreads through the n+ layer towards ground. As the charge spreads, signals are induced in the electrodes, as described in ~\cite{TORNAGO2021165319}. By studying how the signals are shared between the electrodes, the location of the particle interaction can be precisely interpolated with precision much finer than the segmentation of the electrodes.  As AC-LGADs offer a simple fabrication, maintain the precise time resolution of DC-LGADs, and allow fine segmentation, they serve as a strong sensor candidate for 4-dimensional trackers.

The AC-LGAD sensors are sufficiently new that the possibilities for the electrode optimization have just begun to be explored, including variations in the pitch, shape and size of electrodes, and the resistivity of the n+ implant. 
We reported first studies of the performance of prototype AC-LGAD sensors exposed to particle beams in a previous publication~\cite{Apresyan:2020ipp, TORNAGO2021165319}. 
In this paper we expand those studies to include several new detector prototypes produced by Brookhaven National Laboratory (BNL) and the KEK/Tsukuba group in collaboration with Hamamatsu Photonics K.K. (HPK)~\cite{HPKsensorRef} at the Fermilab (FNAL) Test Beam Facility (FTBF)~\cite{FTBF}, and focused on in-depth studies of their performance. 
Significant upgrades were made to the experimental setup that result in greatly improved measurement precision.
Section~\ref{sec:sensor} presents a description of the AC-LGAD sensors, and the experimental setup at the FTBF is described in Section~\ref{sec:setup}. 
Experimental results are presented in Section~\ref{sec:results}. 
These results include measurements of sensor signal properties, measurements of the device efficiency, and measurements of position and time resolution. 
Conclusions and outlook are presented in Section~\ref{sec:conclusions}.

\section{The AC-LGAD sensors}\label{sec:sensor}

We studied four different AC-LGAD sensors, designed and produced by BNL and KEK/Tsukuba with HPK.

\begin{figure}[htp]
\centering
\includegraphics[width=0.40\textwidth]{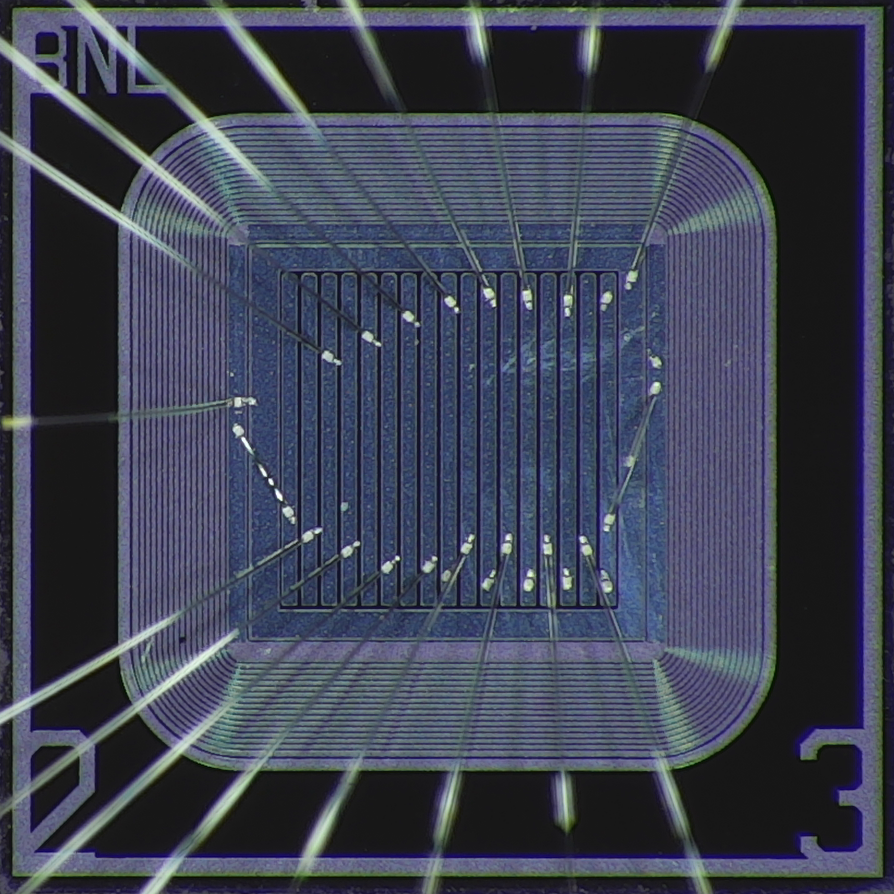}
\includegraphics[width=0.40\textwidth]{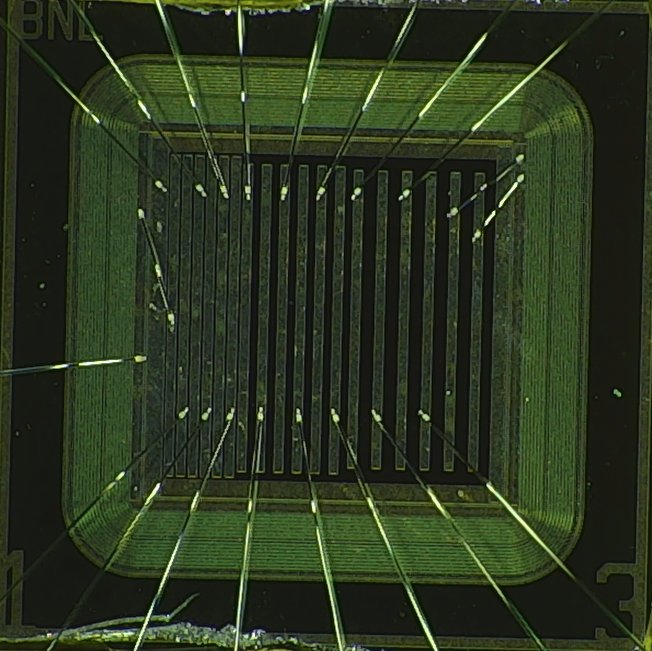}
\caption[Sensor picture]{The BNL manufactured sensors tested at FNAL. BNL 2020 sensor (left) with 100~$\mu$m pitch and 20~$\mu$m gap sizes. BNL 2021 sensor (right) with three pitch variations 100~$\mu$m (narrow), 150~$\mu$m (medium), and 200~$\mu$m (wide).}
\label{fig:BNL2020Pic}
\end{figure}

The first two sensors under study were fabricated at BNL in 2020 and 2021, see Fig.~\ref{fig:BNL2020Pic} for an image of the sensors where the BNL 2020 (BNL 2021) is shown on the left (right) of the figure. 
They were both fabricated using a class-100 silicon processing facility following the process outlined in reference~\cite{ACLGADprocess}.
The sensors come from two different wafers, which underwent a very similar process in the clean room.
The wafers are 4-inch p-type epitaxial wafers, where the 50~$\mu$m-thick epitaxial layer was grown over a 0.5~mm thick low-resistivity substrate. 
The epitaxial layer has a depletion voltage of \SI{120}{\V}.

The first AC-LGAD sensor, shown on the left panel of Fig.~\ref{fig:BNL2020Pic} (BNL 2020), has an active area of about $2\times2$~mm${}^2$, and consists of an array of 17 AC-coupled metal strips surrounded by a DC-connected pad. 
The sensors are \SI{100}{\um} in pitch consisting of \SI{1.7}{\mm} long metallized pads which are \SI{80}{\um} wide with an inter-strip gap of \SI{20}{\um}. 
This sensor has been used in a previous study~\cite{Apresyan:2020ipp}, where more details of its cross section and parameters are given.
The second AC-LGAD sensor, shown on the right panel of Fig.~\ref{fig:BNL2020Pic} (BNL 2021), was cut from a different wafer. 
It has an active area of $3\times3$~mm${}^2$ and consists of three different arrays each with 6 AC-strips that are \SI{80}{\um} wide. 
The three arrays have pitches of 100, 150 and 200 $\mu$m with a strip length of about \SI{2.5}{\mm}, varying slightly between the three arrays to keep the distance from the grounded metal of the DC-contact as large as their gap with neighboring strips.
The two sensors were fabricated with the same thickness of the dielectric of about \SI{100}{\nm} of PECVD silicon nitride, the same dose of $n^+$ implant of about 100 times less than in a standard DC-LGAD, similar breakdown voltages of about \SI{200}{\V}, and the same geometry of the termination.

\begin{figure}[htp]
\centering
\includegraphics[width=0.40\textwidth]{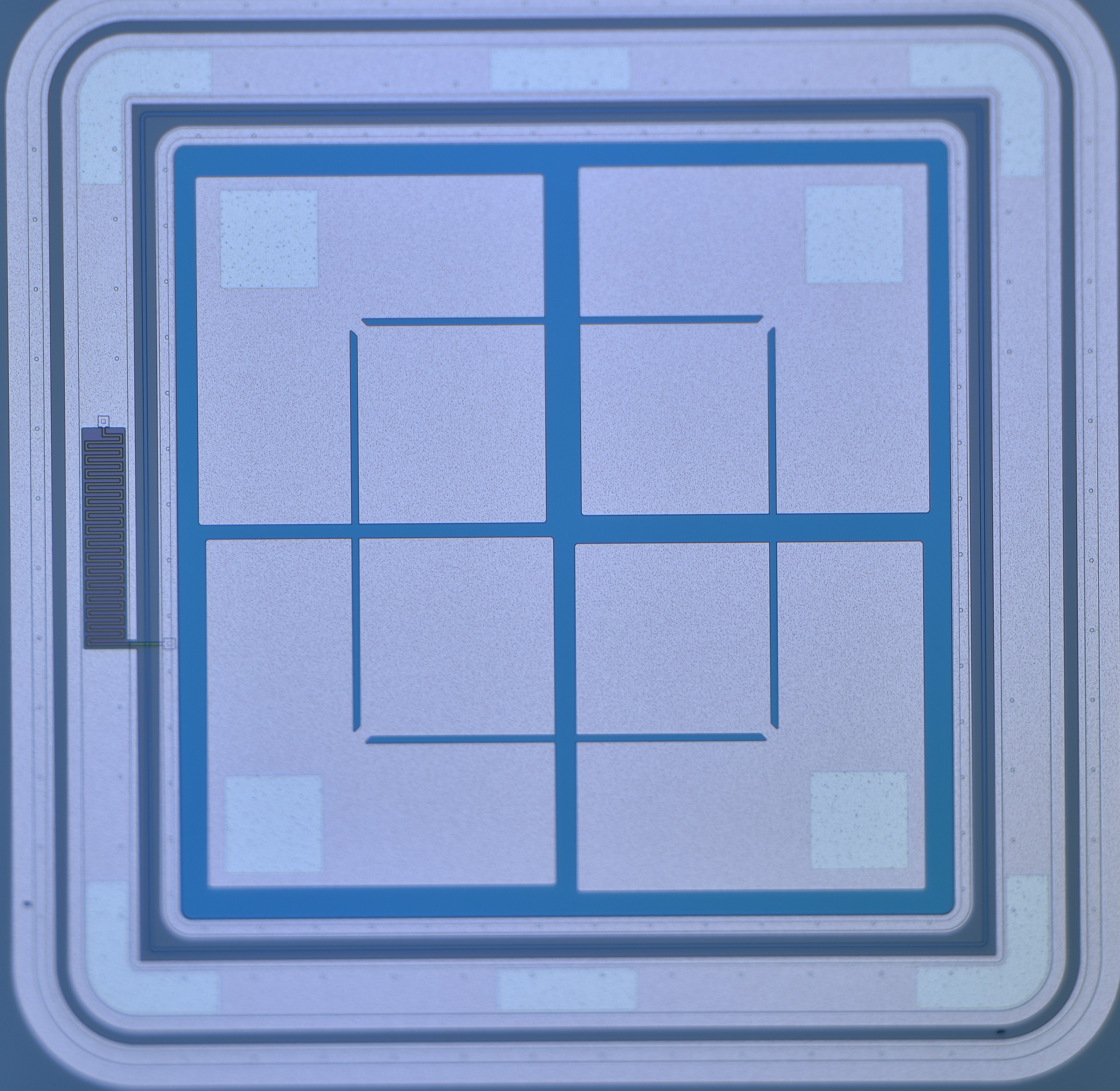}
\caption[Sensor picture]{The HPK manufactured sensor tested at FNAL. The four-pad device with each pad of size $500\times500~\mu$m${}^2$, and interpad gap sizes of 20, 30, 40, and 50 \si{\micro\m}.}
\label{fig:HPK2020Pic}
\end{figure}

The remaining AC-LGAD prototype sensors were developed by KEK/Tsukuba using the HPK layout presented in reference~\cite{HPKsensorRef}. 
Both of them, i.e. the third and fourth sensors, have identical geometry, shown in Fig.~\ref{fig:HPK2020Pic}, where they both have four pad electrodes of size $500\times500$~$\mu$m${}^2$, and feature varying width metalization gaps at each interpad boundary, of 20, 30, 40, and 50 \si{\micro\m}.
Each of the HPK devices has an active thickness of \SI{50}{\um}. 
A variety of $n^+$ and $p^+$ doping concentrations, ranging from A through E and 1 through 3, respectively, were studied in test samples to optimize signal size; the two pad devices discussed here were fabricated with C--2 and B--2.
These differences in the fabrication of each HPK sensor leads to the naming convention used throughout the paper as the HPK C--2 and HPK B--2.
The HPK pad sensors have relatively low $n^+$ resistivity with the aim of using signal sharing to achieve improved position resolution using high-resolution readout digitization. 

\section{The experimental setup at the FNAL Test Beam Facility}\label{sec:setup}

The data presented in this paper were collected at the FNAL test beam facility.
The FTBF supplies a \SI{120}{\GeV} proton beam as well as a silicon tracking telescope to measure the position of each incident proton.

A detailed description of the experimental setup can be found in reference~\cite{Apresyan:2020ipp}, which presents the results of a 2020 beam campaign. Since that time, several key upgrades have been made which greatly enhance the capabilities of the setup. 
These include modifications to the use of the tracking telescope, the oscilloscope used for waveform acquisition, the readout electronics, and the trigger. We describe the upgrades in this section.

\begin{figure}[ht!]
\centering
\includegraphics[width=0.85\textwidth]{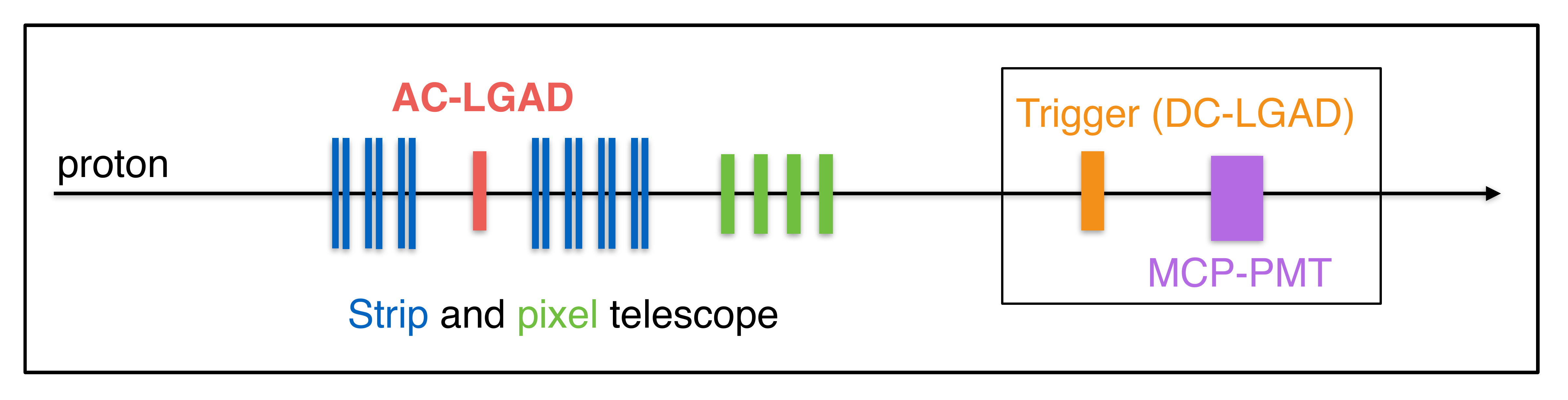}
\caption{A diagram of the AC-LGAD under study and the reference instruments along the beamline.
\label{fig:FTBF_cartoon}}
\end{figure}

\begin{figure}[ht!]
\centering
\includegraphics[width=0.99\textwidth]{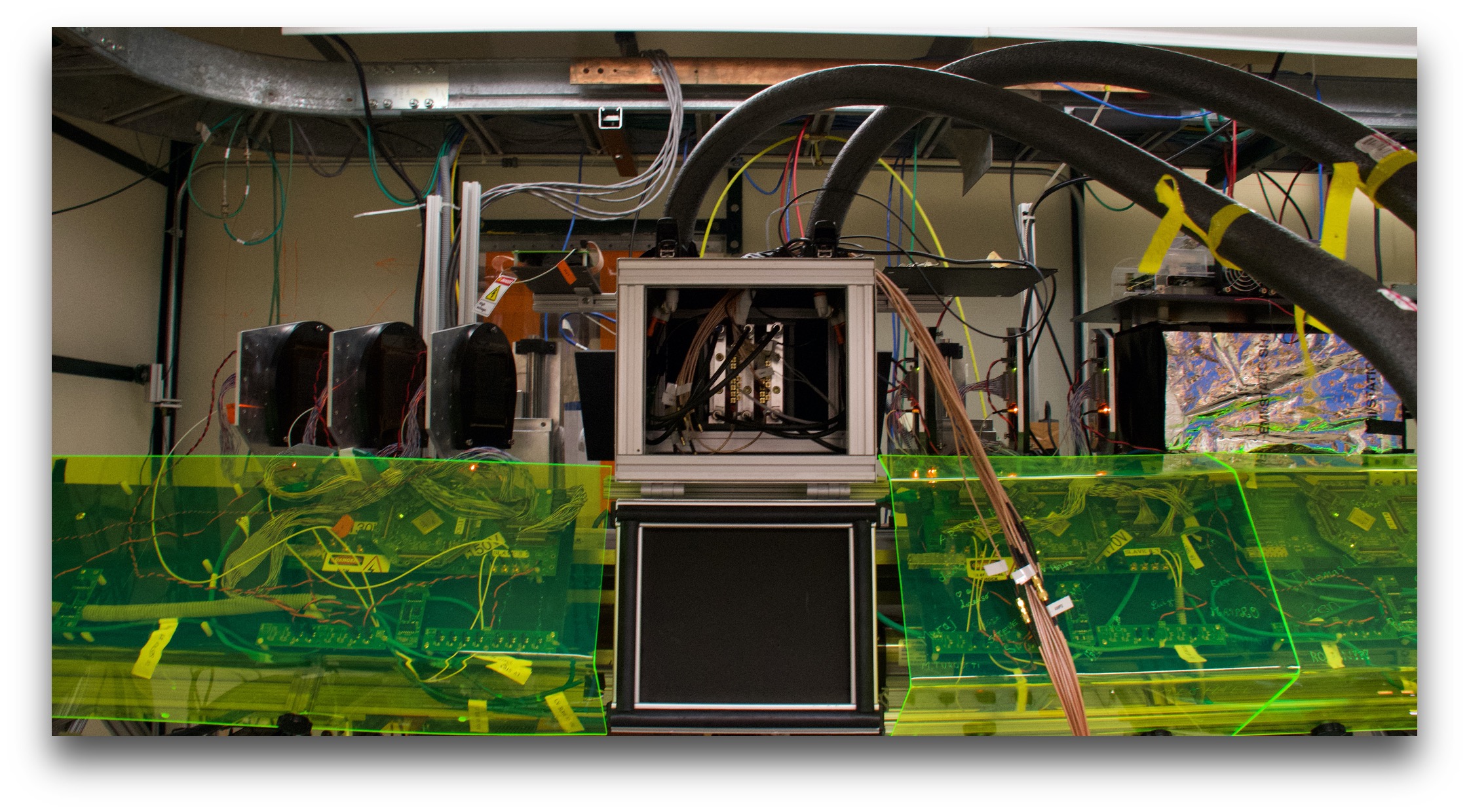}
\caption[FTBF Box]{A photograph of the environmental chamber placed within the FTBF silicon telescope.
\label{fig:FTBF_Box}}
\end{figure}

In the previous setup, the silicon tracker reference resolution was degraded by the extrapolation of the proton track to the location of the AC-LGAD sensor about \SI{2}{\m} downstream. 
The resulting resolution at the sensor location was about \SI{50}{\um} which rendered the measurement insensitive to the AC-LGAD position resolution. 
A new environmental chamber was constructed in order to host the sensors at the center of the telescope, as shown in Fig.~\ref{fig:FTBF_cartoon} and ~\ref{fig:FTBF_Box}. 
At this location, the position resolution of the proton track is in the range of 5--10~\si{\um}, which is comparable to the expected resolution of the AC-LGAD sensors. 
This improvement allows extracting more stringent conclusions about the position resolution of the AC-LGAD sensors. 

The oscilloscope used in the previous campaign (Keysight MSOX92004A) provided only four readout channels, which greatly limited the ability to study the properties of multi-channel clusters. 
For this campaign, a Lecroy Waverunner 8208HD oscilloscope was acquired. 
This oscilloscope features eight readout channels with a bandwidth of \SI{2}{\giga \hertz} and a sampling rate of \SI{10}{GS/s} per channel. 
With the increased channel count, it was possible to capture fully contained clusters within the active area of the sensors, and still provide high quality waveforms for excellent position and time resolutions. 
The AC-LGAD under study typically occupied seven readout channels, while the eighth was used for the Photek micro-channel plate detector (MCP-PMT) as a reference for the proton arrival time. 
The MCP-PMT time resolution for signals from protons has been measured to be about \SI{10}{\ps}. 

The AC-LGAD sensors were mounted on the readout boards developed by Fermilab~\cite{HELLER2021165828} and optimized for LGAD sensors, capable of accommodating 16 independent channels. 
A similar version of this readout board was used in the previous measurement. 
For the current study, the boards used featured a better optimized feedback resistor to improve the signal-to-noise ratio by roughly 70\% at a small cost in signal bandwidth. 
As a result, LGAD sensors read out by the improved 16-channel board now obtain timing performance comparable to the low-noise single-channel UCSC readout board~\cite{Cartiglia201783}. 
The total transimpedance on the 16-channel board after two stages of amplification is roughly \SI{4.3}{\kilo \ohm}, and no additional external amplifiers were used. The gain for LGAD signals is such that each \si{\femto \coulomb} of input charge results in an additional \SI{5}{\milli \volt} in signal amplitude (that is, \SI{100}{\milli\volt} output for a \SI{20}{\femto\coulomb} input).

Additionally, an improved trigger was developed for this campaign, based on a \SI{3}{\milli \m} DC-LGAD sensor mounted on an independent motion stage. 
The trigger DC-LGAD could be aligned with the AC-LGAD under study in order to trigger only a narrow region of interest (ROI). 
Since the AC-LGADs are small relative to the width of the beam, the ROI trigger greatly increased the dataset purity and the acquisition rate of useful events. 
This improvement enabled a much larger set of sensors to be studied in a variety of operating conditions within the limited period of available beam time.

Since the position resolution of the tracker telescope measurement at the AC-LGAD location is now comparable to the expected resolution of the AC-LGAD sensors, the calibration of the tracker telescope is extremely important. 
Several methods were used to estimate the tracker position resolution. 
First, the track reconstruction algorithm based on the Kalman filter provides an estimate of the impact parameter resolution for each track based on the compatibility of the track fit with the observed hits. 
This yields an estimated resolution of roughly \SI{5}{\um}. 
Second, the resolution was estimated by studying the smearing of the efficiency turn-on curve at the edge of a DC-LGAD sensor. 
Assuming the response of the LGAD at the pad edge is a step function with a turn-on sharper than \SIrange{1}{2}{\um} microns, then the observed smearing can be attributed to the tracker resolution and extracted by a fit to an error function.
Finally, the resolution was also estimated using strip sensors with known pitch using binary readout.
In a binary readout scheme, the resolution is simply the pitch divided by $\sqrt{12}$. 
Any enhancement of the resolution beyond this value is attributed as the tracker impact parameter resolution. 
In both the DC-LGAD edge fit and the binary readout methods, the resolution obtained is in the range of \SIrange{10}{12}{\um}, somewhat at tension with the \SI{5}{\um} result obtained from the Kalman filter. 
Since the Kalman filter resolution does not include contributions from the alignment with the device under test, we take its estimate as a lower bound. 
Conversely, since the two other methods may contain additional systematic errors, we take their estimates as upper bounds on the resolution. 
As result, we limit the tracker resolution to a range of roughly 5--10 \si{\um} and consider this range to interpret the observed AC-LGAD performance. For the AC-LGAD resolutions presented in all plots and figures in this paper, we subtract a contribution of \SI{6}{\micro\m}, representing a conservative choice for the tracker resolution.

\section{Experimental results}\label{sec:results} 

\begin{figure}[htp]
\centering
\includegraphics[width=0.55\textwidth]{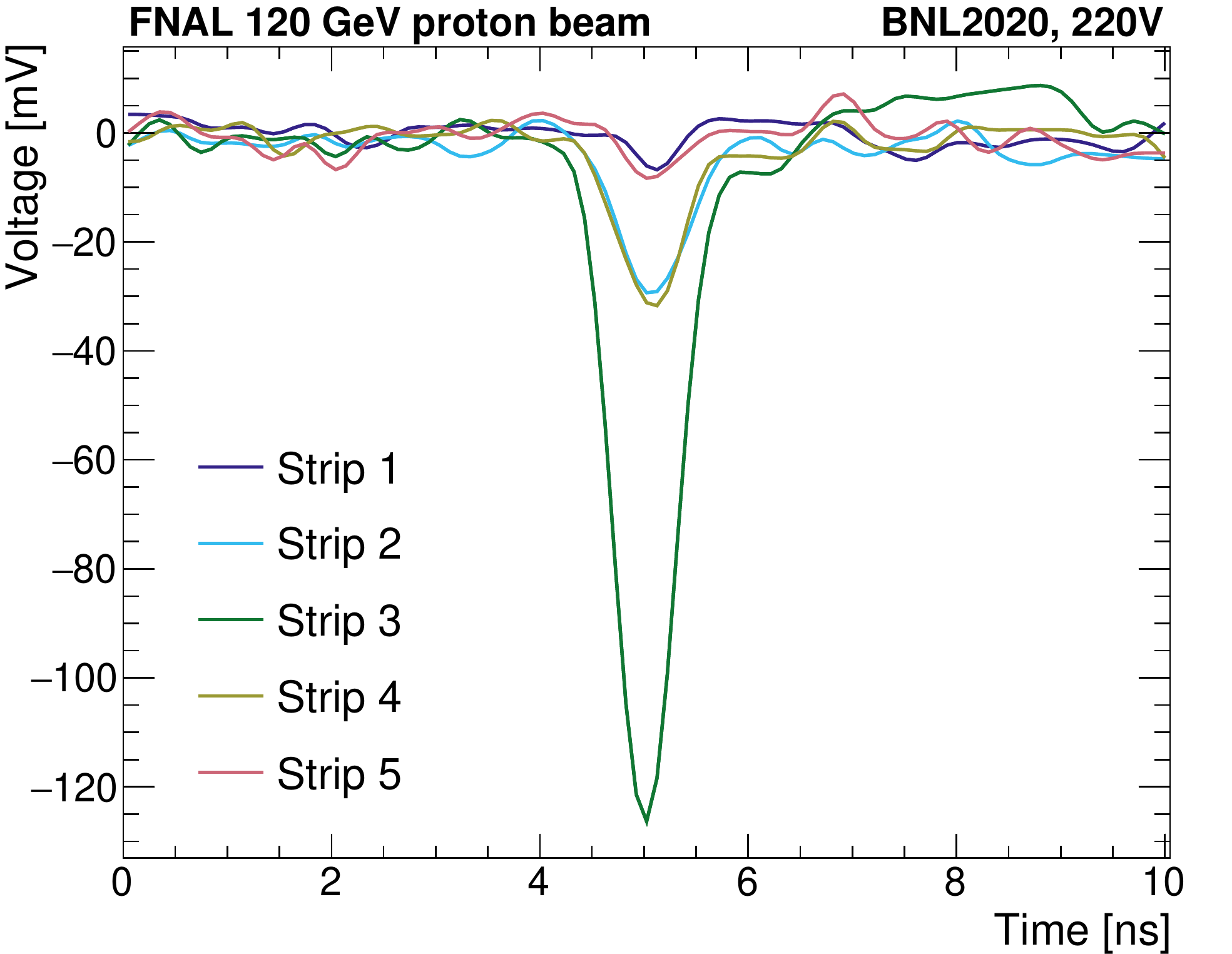}
\includegraphics[width=0.44\textwidth]{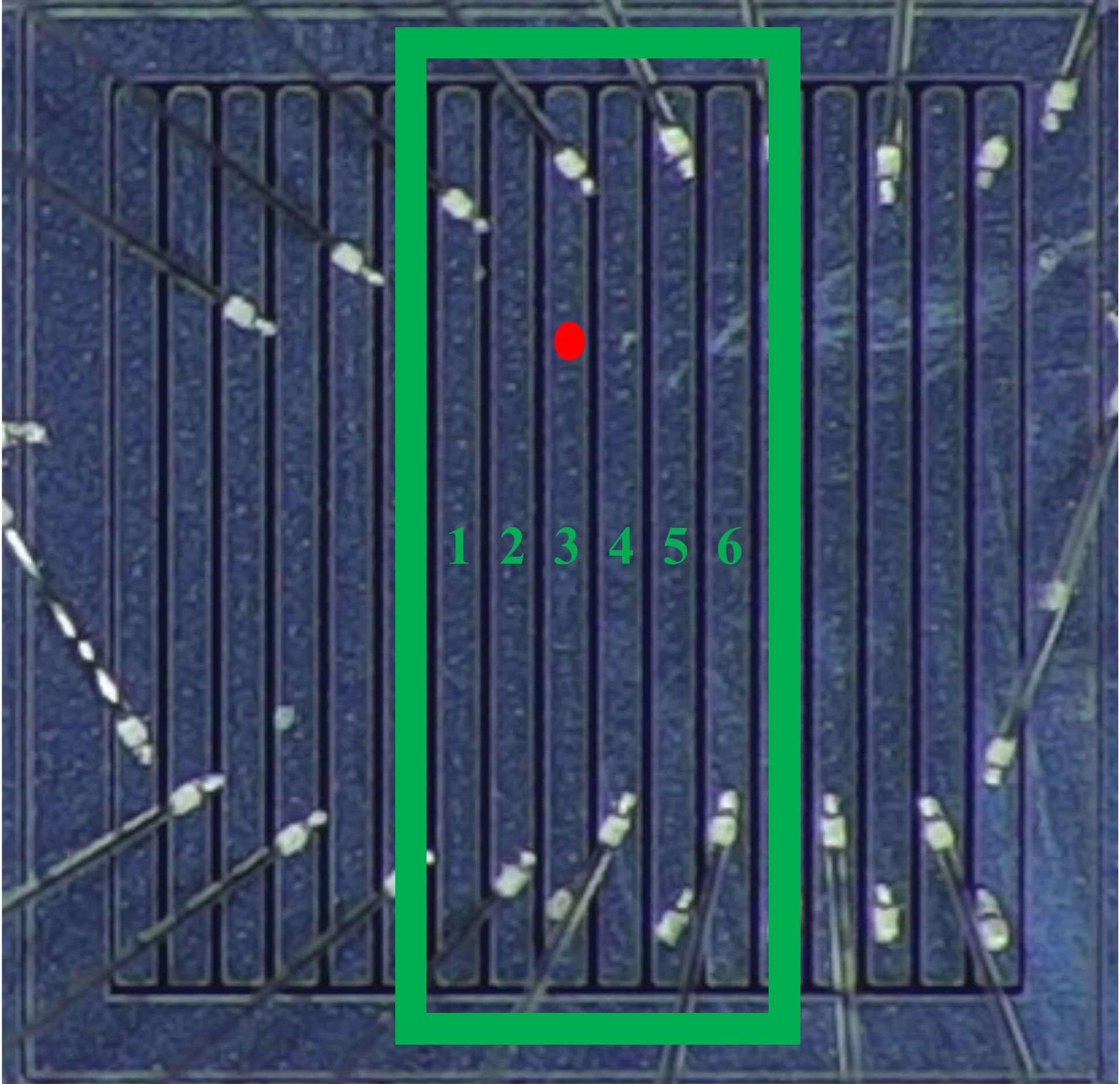}
\caption[Waveform]{Example waveforms for the BNL 2020 strip sensor (left) showing the measured pulse shape and size for all channels in an event with a proton impacting the third readout strip. A labeled photograph of the six readout strips is shown on the right with a red dot indicating the location of the proton hit for the example waveforms shown on the left.
\label{fig:WavePlot}}
\end{figure}

Digitized waveforms of the analog signals were recorded by the oscilloscope for each sensor tested, as described in Section~\ref{sec:setup}. We analyze the waveforms to measure the amplitude and arrival time of pulses in each individual channel. Waveforms for a typical example event produced by the BNL 2020 strip sensor are shown in Fig.~\ref{fig:WavePlot}.
In general, the waveforms from each sensor have similar shapes, which allows reconstructing the amplitude and time for each charged particle hit using the same algorithm for all sensors.

Events were selected based on two types of requirements. 
First, only events with high-quality tracks and MCP-PMT hits are considered, to ensure reliable references for the proton impact parameter and arrival time. 
We further require that the track points through the interior of the readout region of each sensor, to exclude clusters that are only partially reconstructed at the edges. 
For example, for the BNL 2020 strip sensor shown in Fig.~\ref{fig:WavePlot}, only events with tracks pointing between strips 2--5 are considered for the analysis. 
Then, we define two amplitude thresholds that are used to select good events as well as define channels that can be used for the reconstruction steps described later in this section. 
A primary threshold ranging from 30--40~\si{\mV} is used to select channels consistent with a direct proton hit, and a secondary threshold of 10--20~\si{\mV} identifies secondary signals that are not consistent with noise fluctuations. 
A range of thresholds were used, optimized for each sensor, and the same threshold was subsequently used for each sensor for all of the measurements.
Only events with a hit above the primary threshold are considered, and the primary strip is defined as the strip with the largest amplitude.




Each sensor was studied at various bias voltages to determine an operating point  that optimized the measured time resolution.
The preferred bias voltage values selected for the results discussed in the remaining sections are -220~V, -285~V, -180~V, and -230~V for the BNL 2020, BNL 2021, HPK C--2, and HPK B--2 sensors respectively. 



\subsection{Signal properties}

After defining an appropriate event selection, we study the signal properties beginning with the size of the signal amplitudes in the primary channel, as well as the neighboring channels. 
Subsequently we combine information from multiple channels to reconstruct the position and time of each proton hit. 

\begin{figure}
    \centering
    \includegraphics[width=0.49\textwidth]{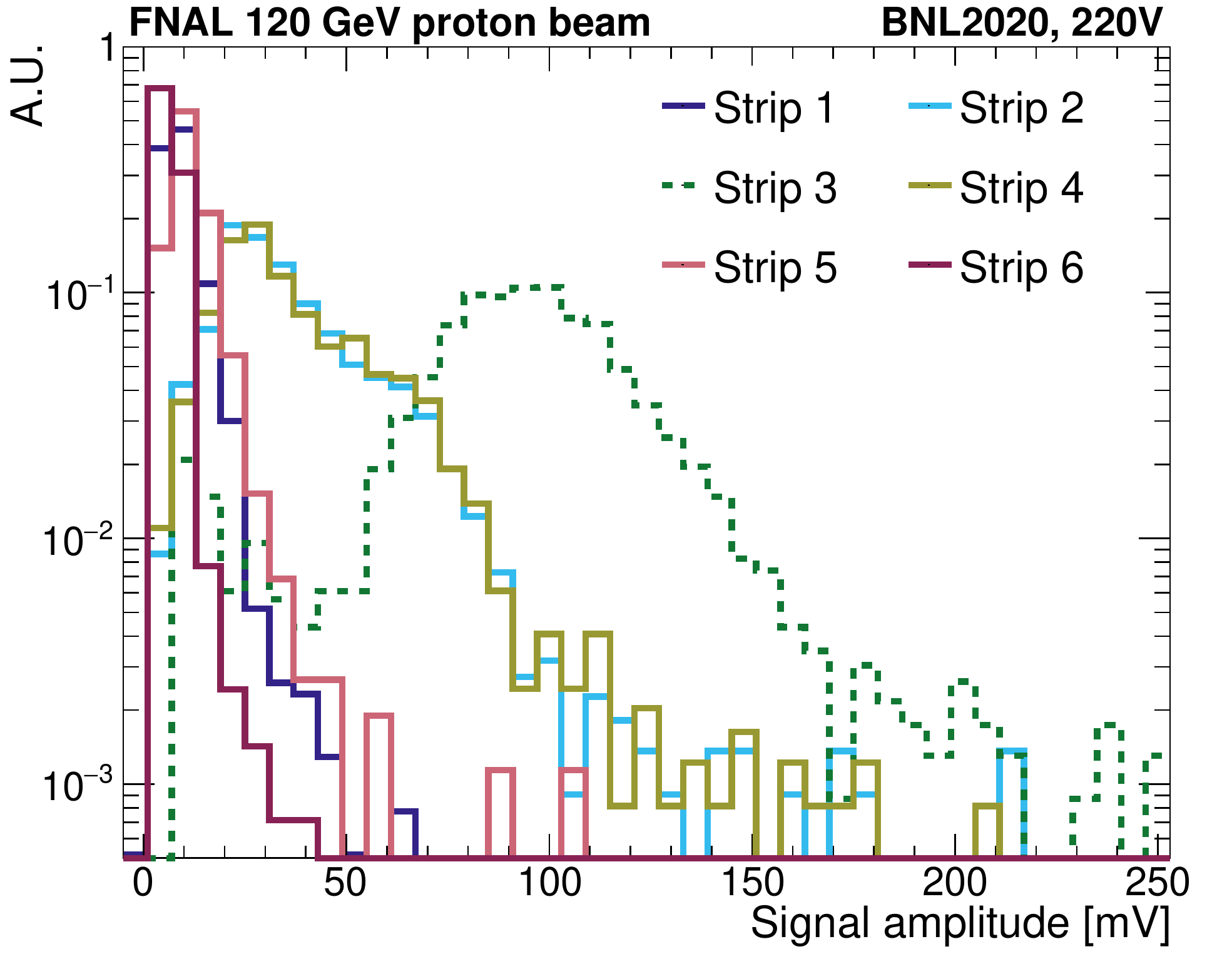}
    \includegraphics[width=0.49\textwidth]{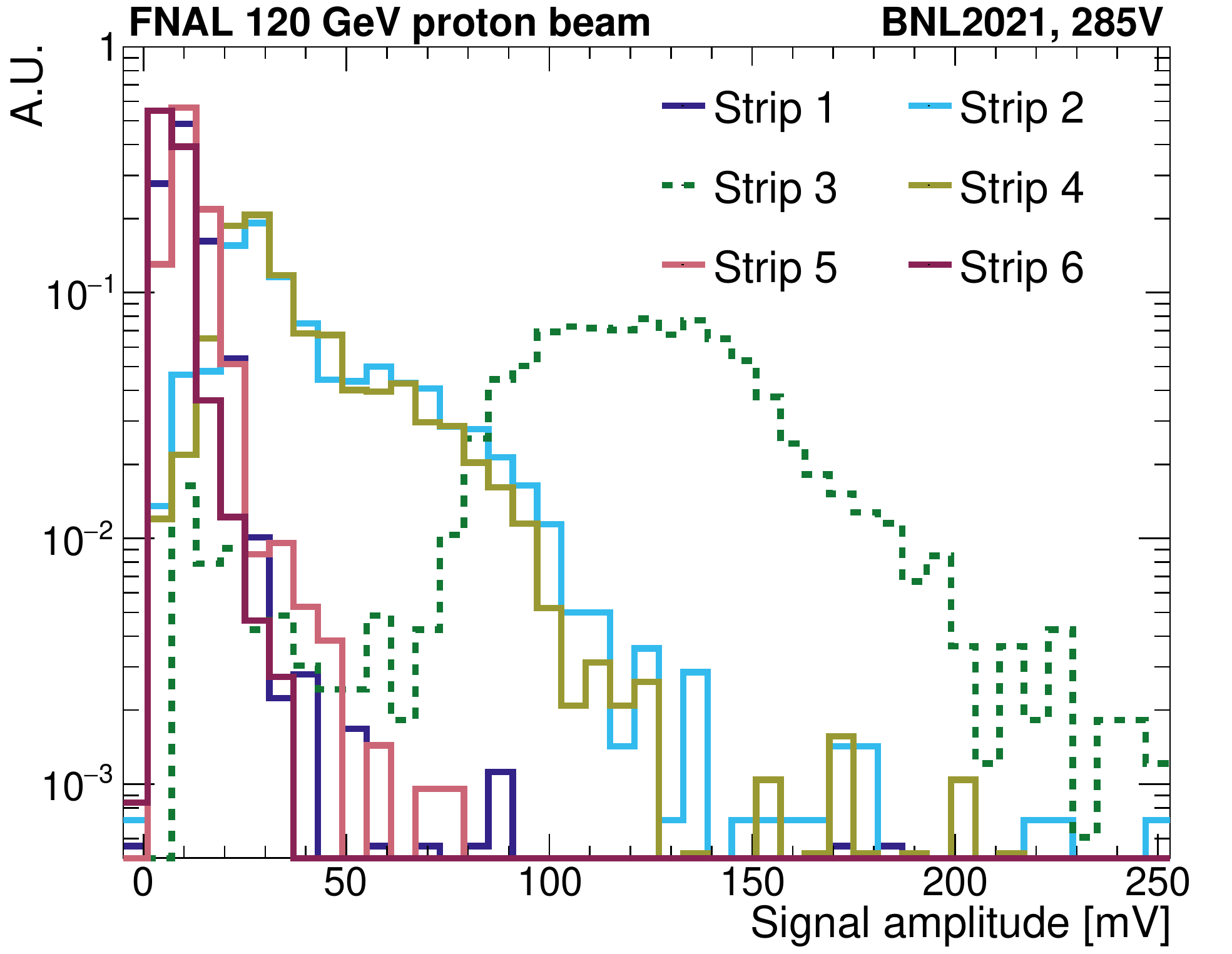}
    \caption{
    Measured amplitudes for all channels in events where the proton hits Strip 3 for the BNL 2020 (left) and BNL 2021 medium pitch (right) sensors.
    Each curve is normalized to unit area.
    }
    \label{fig:AmpChSh_HPKs}
\end{figure}

The typical signal amplitudes are a  critical property for the performance of each sensor.
For the BNL 2020 and BNL 2021 medium pitch sensors, Fig.~\ref{fig:AmpChSh_HPKs} shows the signal amplitudes for events where Strip 3 is the primary channel, following the numbering convention in Fig.~\ref{fig:WavePlot}.
Due to the signal sharing properties of the AC-LGADs, we observe signals with amplitudes well above noise for secondary channels that did not have a direct impact from the proton. 
The size of the signal amplitude decreases for strips that are increasingly further away from the primary strip.


\begin{figure}[htp]
    \centering
    \includegraphics[width=0.49\textwidth]{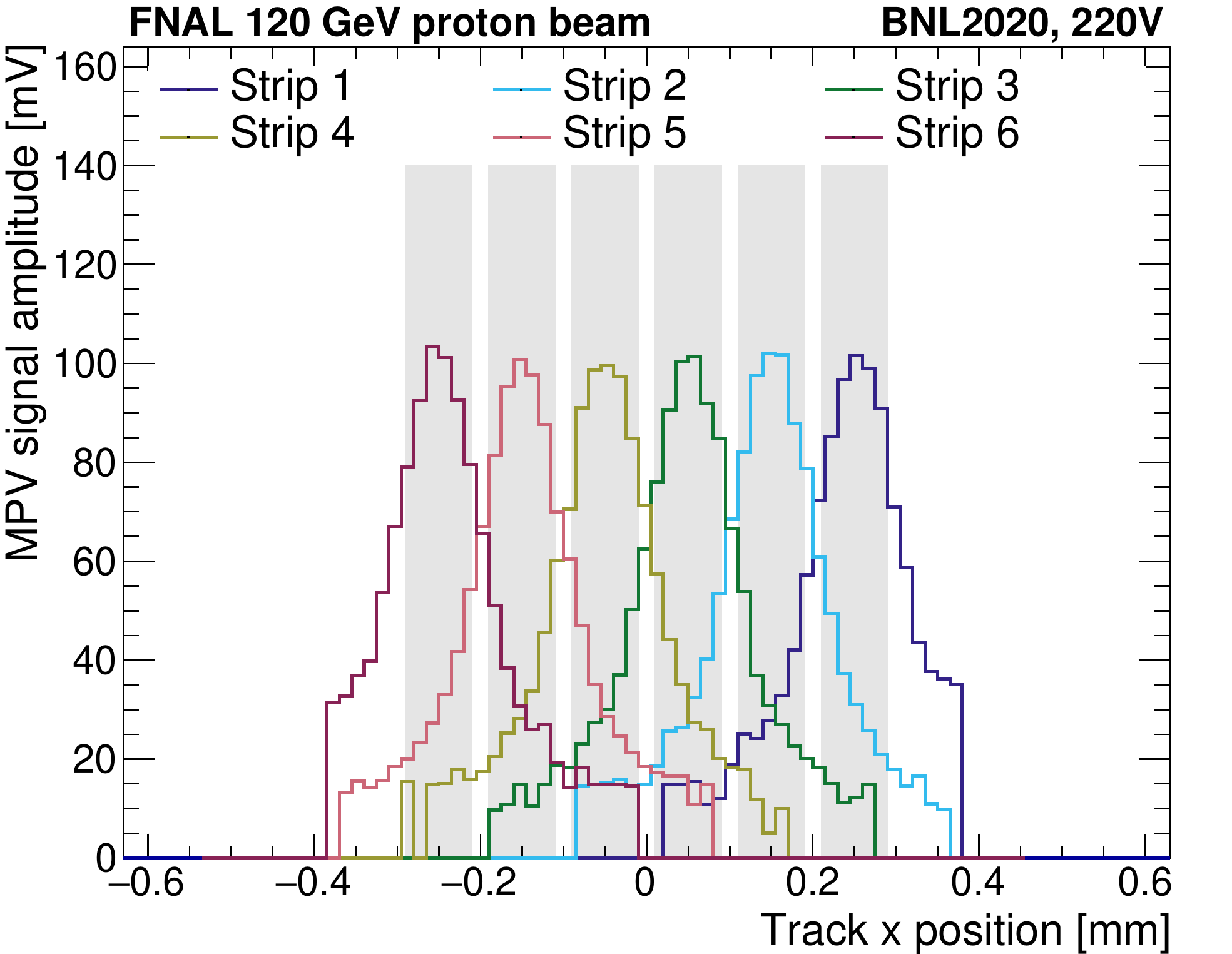}
    \includegraphics[width=0.49\textwidth]{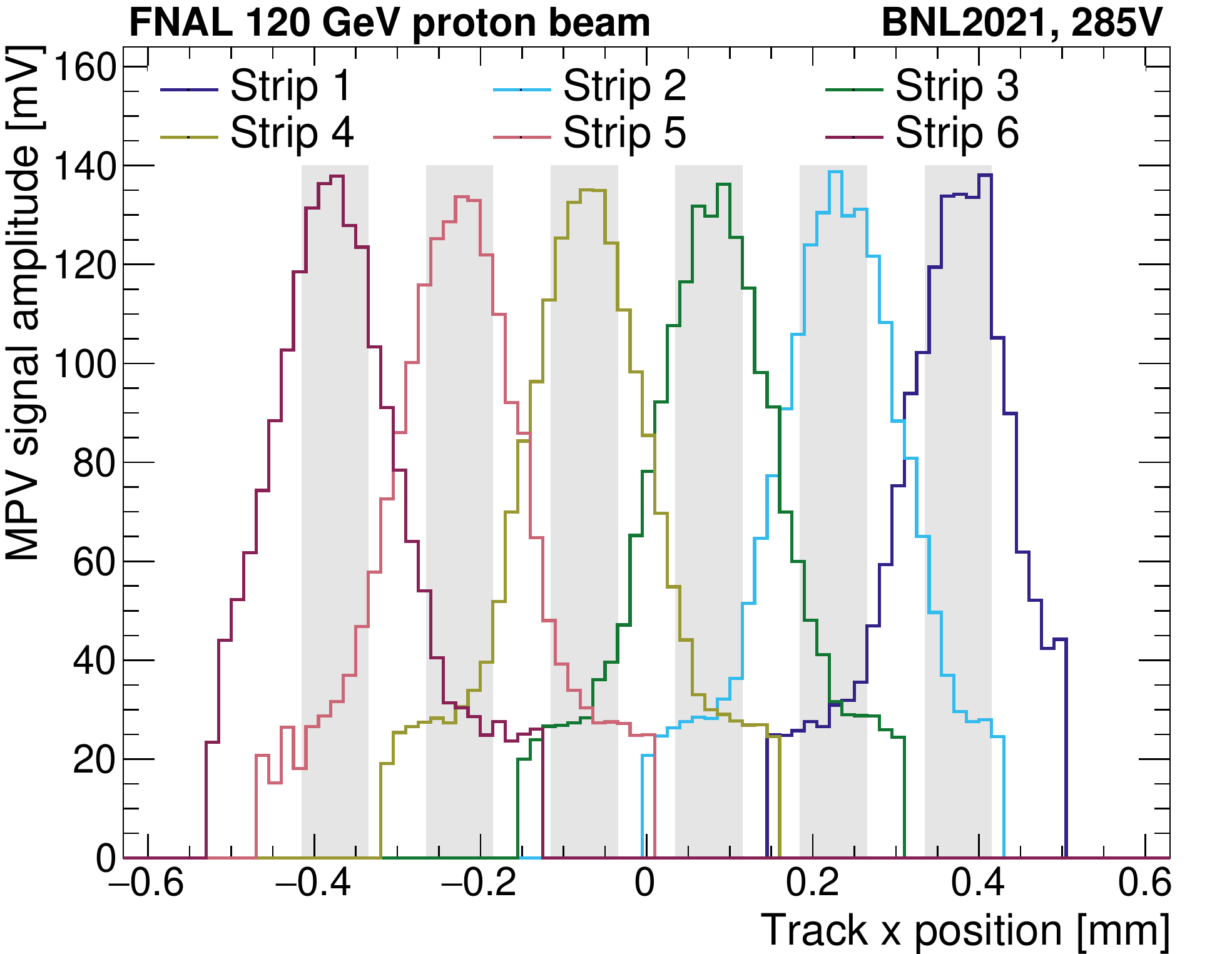}
    \caption{Most probable values of amplitudes as functions of proton track hit position. Each curve shows the MPV value for each strip individually for the BNL 2020 (left) and BNL 2021 medium pitch (right) sensors. The grey area indicates the metallized regions on the sensor surface.
    \label{fig:ampVsXBNL}}
\end{figure}

\begin{figure}
    \centering
    \includegraphics[width=0.49\textwidth]{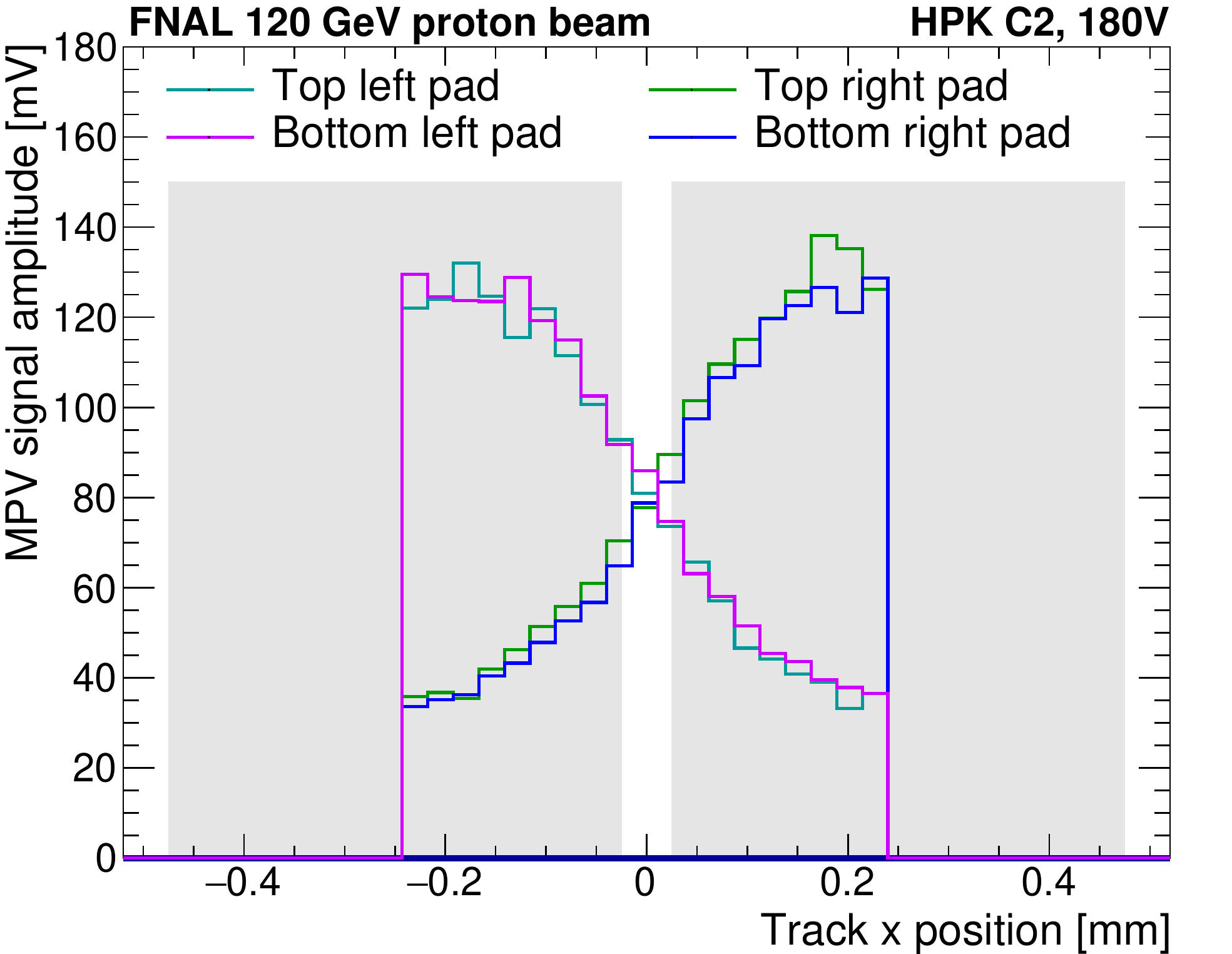}
    \includegraphics[width=0.49\textwidth]{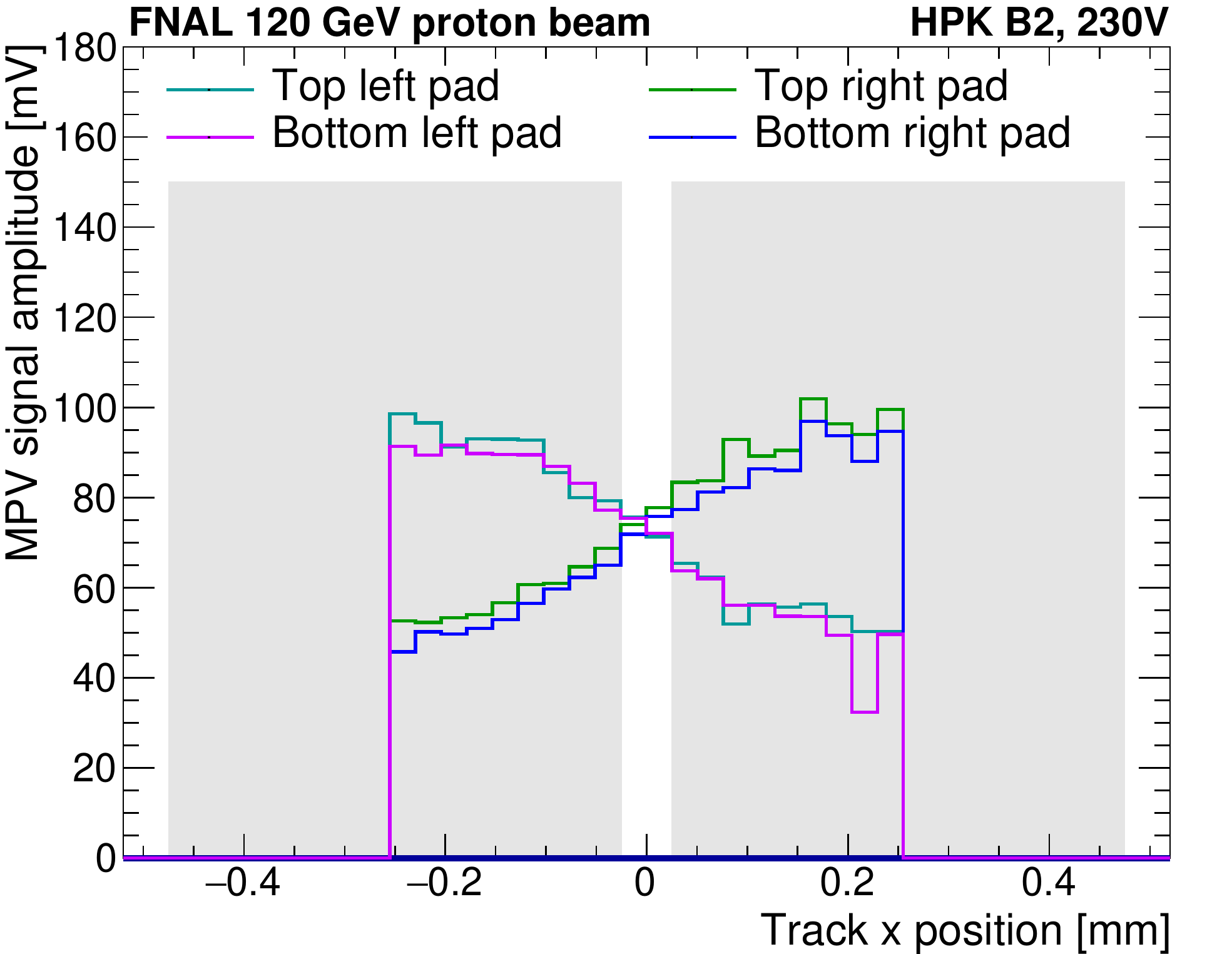}
    \caption{
    Most probable values of amplitudes as functions of the proton track hit position.
    Each curve shows the MPV value for each pad individually for the HPK C--2 (left) and HPK B--2 (right) sensors.
    The amplitudes as a function of track y position shows similar results.
    The grey area indicates the metallized regions on the sensor surface.
    }
    \label{fig:TotalAmp_HPKp}
\end{figure}

Next, we studied the signal properties across the surface of the sensor. In bins of the proton impact parameter, $x$, we fit the signal amplitude histogram to the convolution of a Landau distribution and a Gaussian function to extract the most probable value (MPV) of the amplitude. Figures~\ref{fig:ampVsXBNL}~and~\ref{fig:TotalAmp_HPKp} show the MPV signal amplitudes as a function of $x$ for each channel in the BNL strip and HPK pad sensors, respectively. 
The shaded gray regions indicate the regions covered by the metallized AC contacts.
As seen in Fig.~\ref{fig:ampVsXBNL}, the BNL 2020 and 2021 strips show large signals in particular for channels with a direct proton hit.
From these plots, we see signal sharing that extends to 2-3 strips away from the primary strip before reaching the noise floor.
Figure~\ref{fig:TotalAmp_HPKp} shows the MPV amplitudes for the HPK C--2 and B--2 sensors as a function of $x$ only, while the results are similar for the y direction.
The HPK C--2 and B--2 sensors have small variations in metallic gap sizes across all four gaps in the sensor, but these variations had no impact on the observed performance.
It can also be seen in Fig.~\ref{fig:TotalAmp_HPKp} the impact of the varying resistivity in the HPK C--2 and B--2 sensors. 
The high resistivity C--2 sensor produces larger signals in the primary channel, but the signals decrease more rapidly with distance, resulting in smaller signals in the neighboring channel. 
By contrast, the HPK B--2 sensor shows a flatter distribution with smaller difference between the primary and secondary signals. 
The more equitable sharing of signals in B--2 is a consequence of the larger signal sharing scale associated with a lower resistivity surface layer. 
For future applications, the resistivity should be carefully tuned to provide a signal-sharing distance appropriate for the pitch of the device under consideration.

\begin{figure}[htp]
\centering
\includegraphics[width=0.49\textwidth]{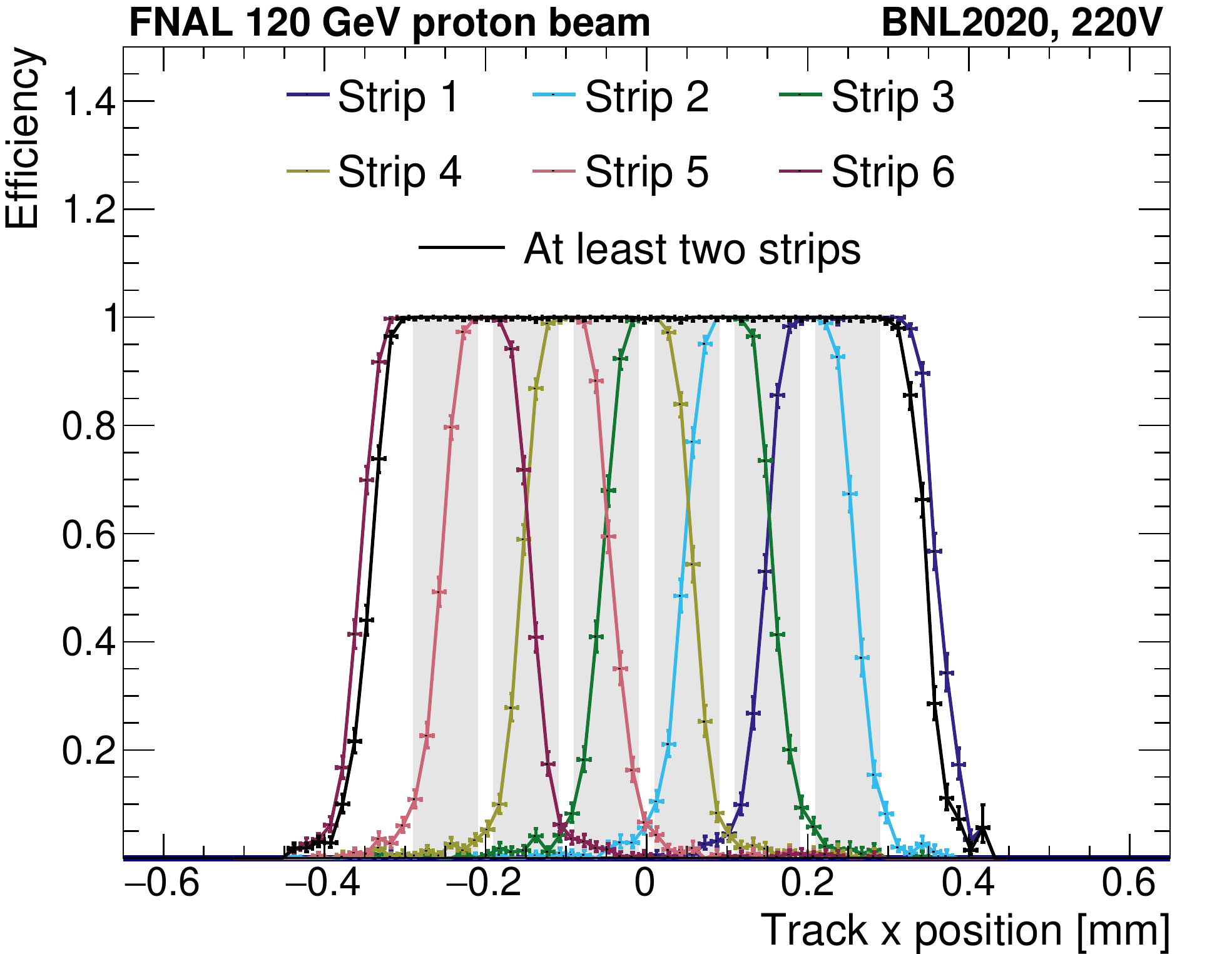}
\includegraphics[width=0.49\textwidth]{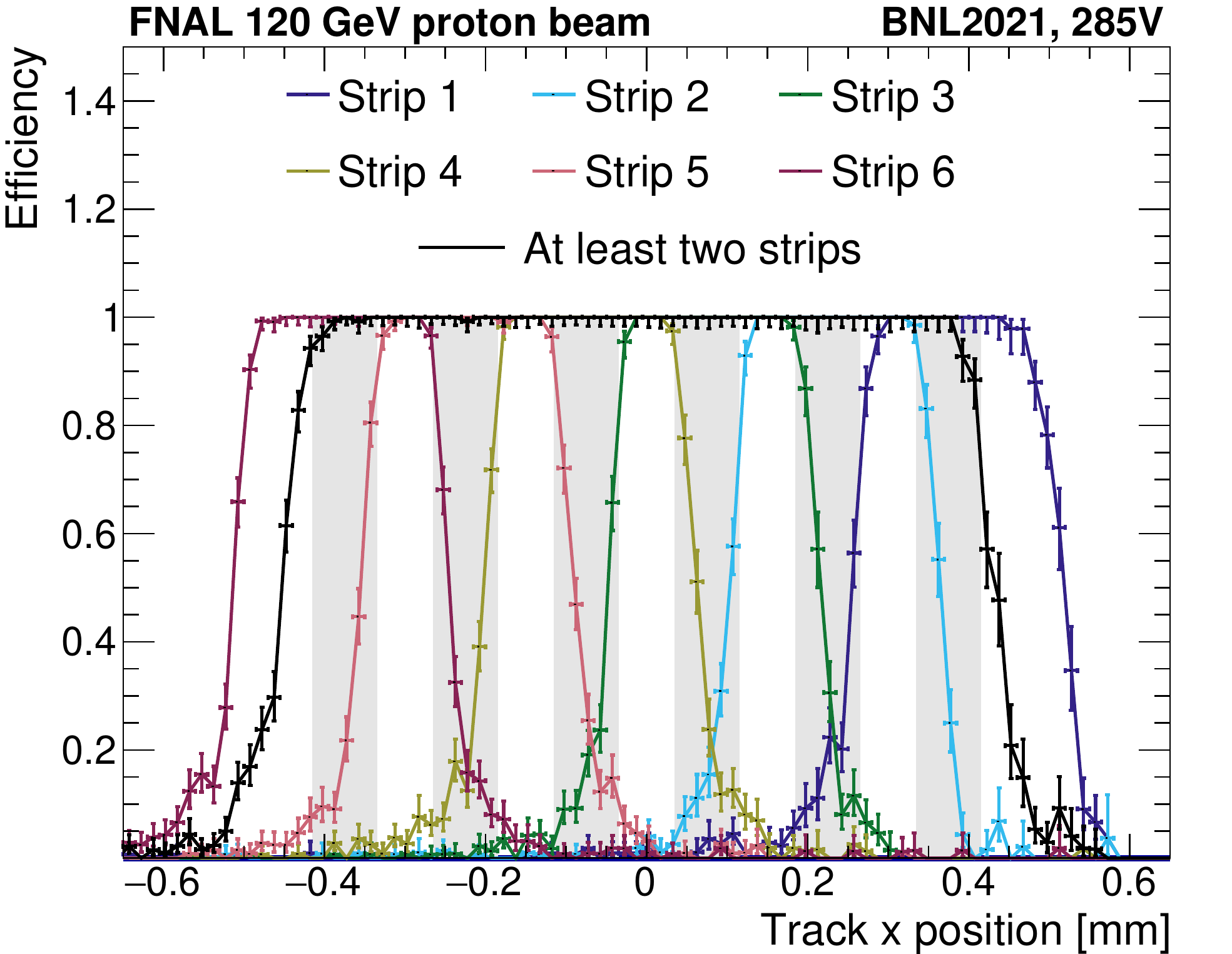}
\caption{Response efficiency of the BNL 2020 (left) and BNL 2021 medium pitch (right) sensors for signal thresholds of \SI{30}{\mV} and \SI{40}{\mV}, respectively.
The black curve corresponds to the logical ``OR'' of all channels, with the additional requirement that there is a second strip with an amplitude above \SI{10}{\mV} (BNL 2020) or \SI{20}{\mV} (BNL 2021).
This curve represents the efficiency of the requirements to perform a two strip position reconstruction.
The grey area indicates the metallized regions on the sensor surface.
\label{fig:effBNLStrip}}
\end{figure}

We also measure the detection efficiency of each sensor, defined as the number of events passing the amplitude requirement discussed above divided by the number of events for which a track was reconstructed and whose hit position is within the active region of the sensor. 
We observed that all sensors were fully efficient across the readout regions for the \SI{120}{\GeV} protons.
Fig.~\ref{fig:effBNLStrip} shows the efficiency of the primary threshold for each strip in the BNL sensors as well as the efficiency for a second strip to additionally pass the secondary threshold. Thanks to the signal sharing, efficiency of 100\% is maintained even in the gaps between pairs of strips. The efficiency to find a primary and secondary channel is also near 100\%, which allows reconstruction of the proton position based on interpolation between the two hit strips, as discussed in the following section.

%

\subsection{Position reconstruction}

The spatial resolution of the AC-LGADs can be significantly improved with respect to binary readout by utilizing the signal sharing property of the sensors. 
To achieve the improved position resolution, we interpolate the position between two strips using the signal amplitude information from multiple neighboring strips.
We present two different methods for achieving this interpolation. 
The first method estimates the proton position using the ratio of signal amplitudes in the primary and secondary strips.
The second method uses a machine learning technique that incorporates all measured signal amplitudes into a neural network trained to reconstruct the track hit position from the telescope.
In both methods the measurements of the position resolution are limited by the performance of the tracker reference, 5--10~\si{\micro\m} resolution, as discussed in the following.

\subsubsection{Amplitude ratio method}

In this method, we consider only the signal amplitudes from the leading and subleading channels.
For the strip sensors, the subleading channel is always on the left or right side of the maximum amplitude channel.
For the pad sensors, there can be multiple pads adjacent to the primary pad with comparable subleading signal amplitudes. The horizontal and vertical position estimates are performed independently, relying on the signals in the horizontal and vertical neighbors of the primary pad.

\begin{figure}[htp]
    \centering
    \includegraphics[width=0.49\textwidth]{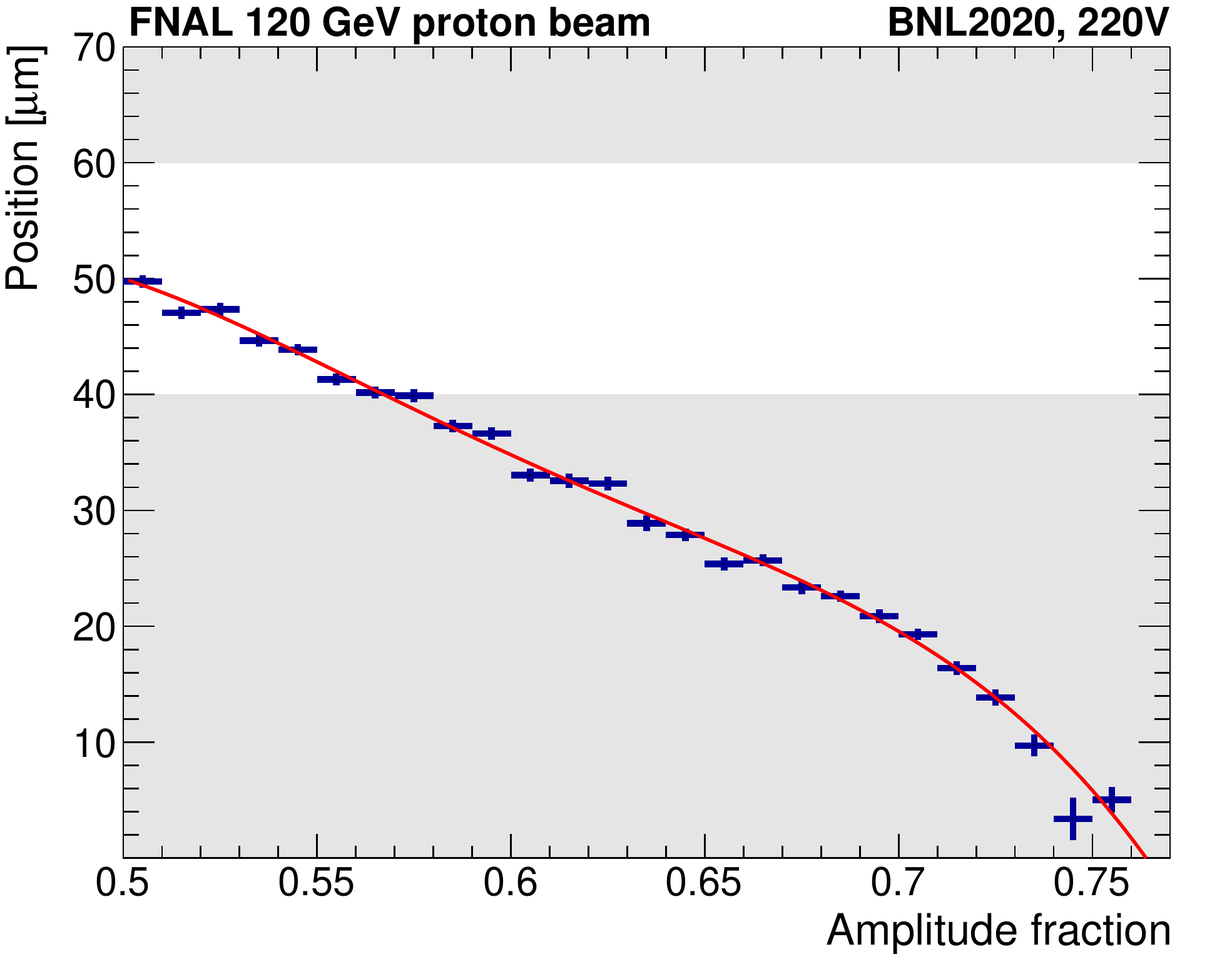}
    \includegraphics[width=0.49\textwidth]{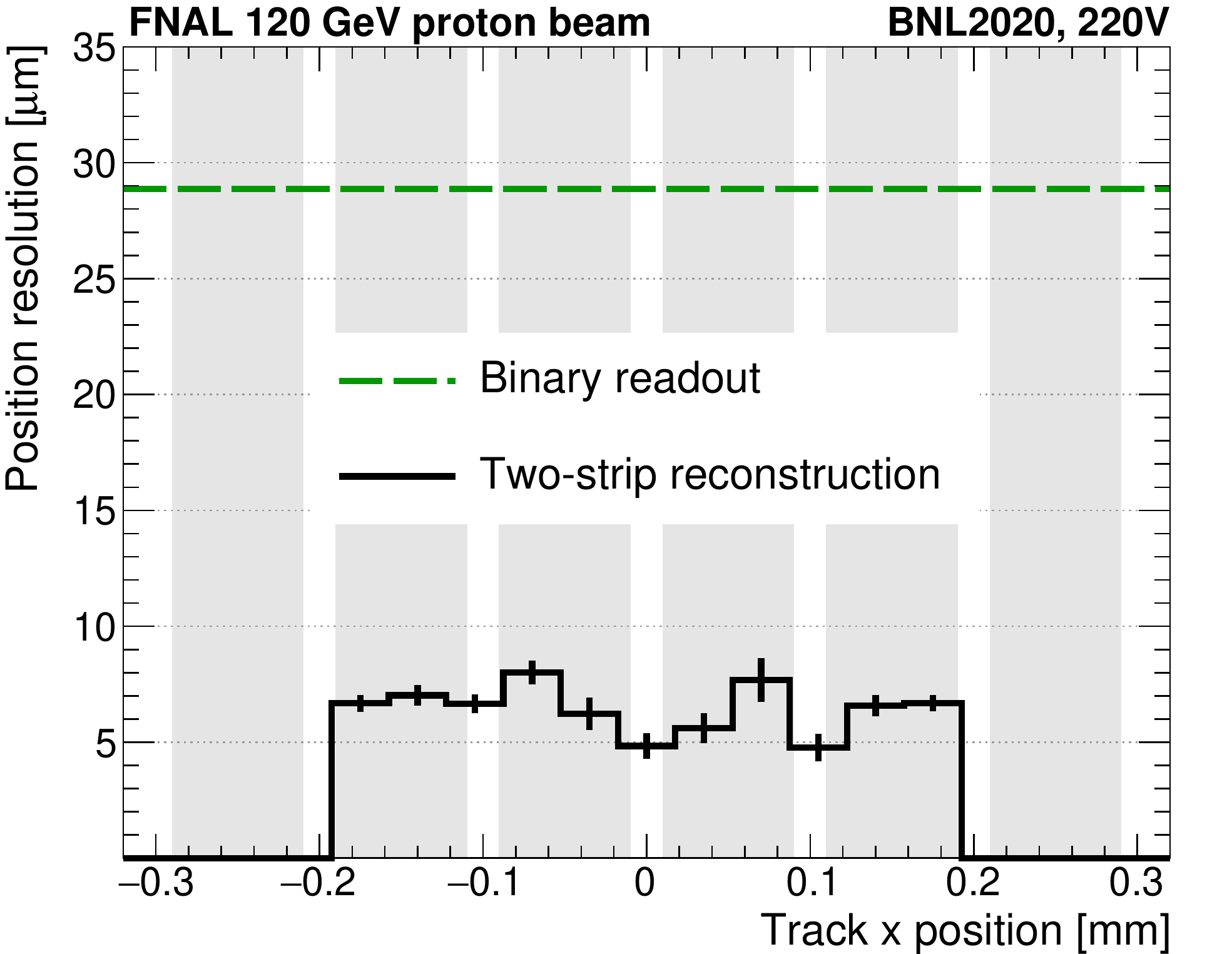}
    \caption{
    Proton track hit position with respect to the center of the primary strip as a function of the amplitude fraction in the BNL 2020 sensor (left). Position resolution as a function of track position for the same sensor (right). The grey area indicates the metallized regions on the sensor surface. Errors represent the statistical uncertainty.
    \label{fig:posFitBNL2020}}
\end{figure}

To perform the position reconstruction, we first define in each event the amplitude fraction, $a_1/(a_1+a_2)$,  where $a_1$ and $a_2$ are the amplitudes of the leading and subleading channels. Then, for each sensor, we study the dependence of the amplitude fraction as a function of the proton track hit position. The mean track position in bins of the amplitude fraction is shown in Fig.~\ref{fig:posFitBNL2020} (left) for the case of the BNL 2020 sensor, over the range of one-half pitch from the center of the primary channel to the boundary with the subleading channel. These distributions are then fit with up to a $5^{\rm{th}}$ degree polynomial. Finally, the proton position in each event is estimated by finding the amplitude fraction and evaluating  the fit polynomial at that fraction to find the distance from the center of the primary channel.

The signal sharing polynomials were obtained using only the data at nominal voltage for each sensor, though it was observed that the relationship is not strongly dependent on the bias voltage. 

Figure~\ref{fig:posFitBNL2020} (right) show the performance of the simple algorithm across the surface of the BNL 2020 sensor. The position resolution of the simple two-strip reconstruction reaches about \SI{6}{\micro\m}, after subtracting in quadrature the mean expected contribution from the tracker resolution of \SI{6}{\micro\m}.
This value represents approximately a factor of 6 improvement from the resolution expected from binary readout, $\rm{pitch}/\sqrt{12}$.

Similar results are obtained using all of the BNL 2021 strip pitch variations. Since the BNL 2020 and 2021 sensors all reach position resolutions better than the tracker reference, it is difficult to distinguish their performances. The limits obtained on their position resolutions are shown in Table~\ref{table:SignalProp}. The observed differences are attributed to systematic variation in the tracker performance and alignment quality for different datasets, rather than differences intrinsic to the sensors.






\begin{figure}[htp]
    \centering
    \includegraphics[width=0.49\textwidth]{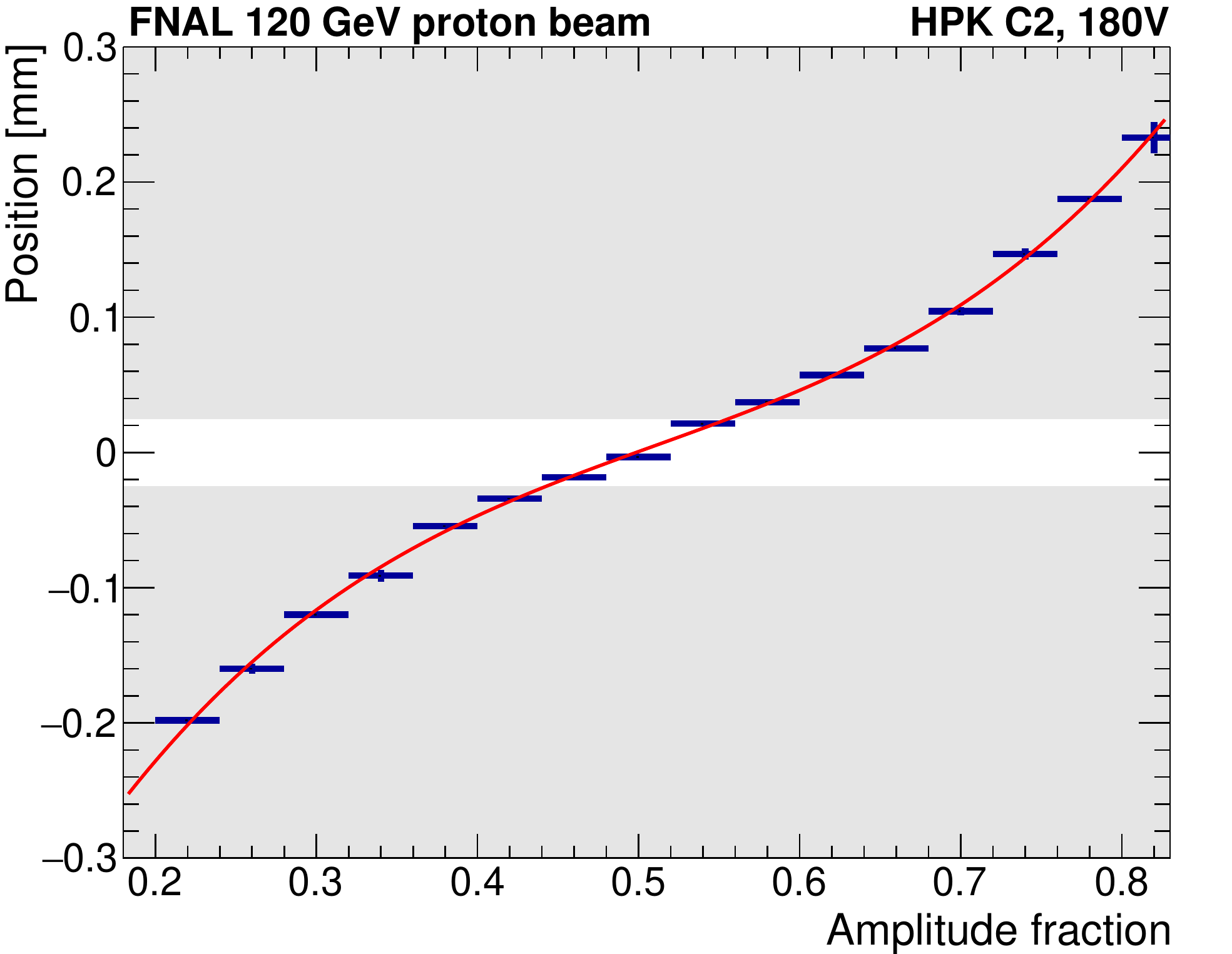}
    \includegraphics[width=0.49\textwidth]{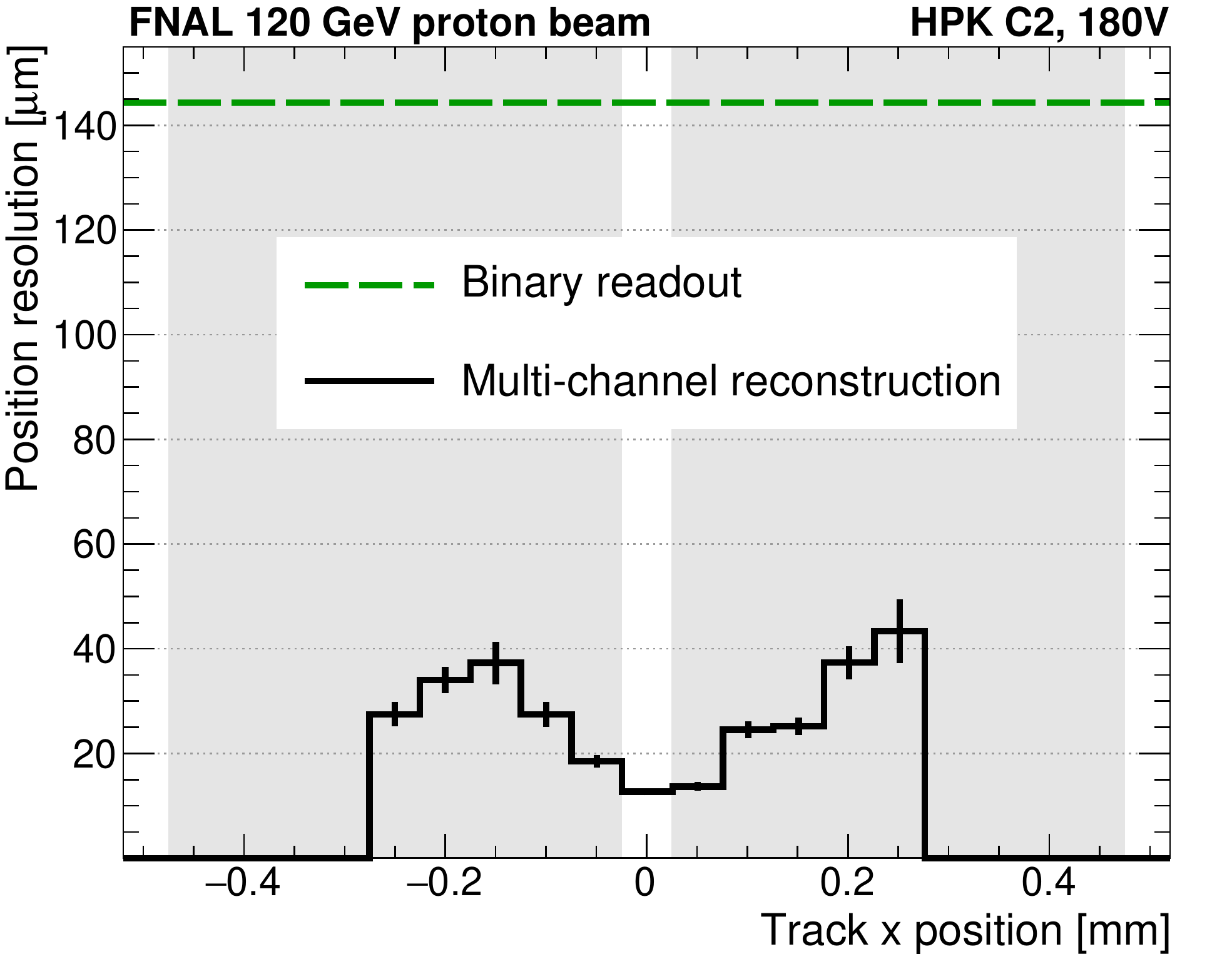}
    \caption{Proton track hit position as a function of the amplitude fraction in the upper right pad over the total amplitude in the upper half of the HPK C--2 pad sensor (left). Position resolution as a function of track position for the same sensor (right). The grey area indicates the metallized regions on the sensor surface. Errors represent the statistical uncertainty on the fits.
    \label{fig:PositionFit_HPKC2p}}
\end{figure}

The amplitude fraction model and the observed position resolution for the HPK C--2 pad array are shown in Figure~\ref{fig:PositionFit_HPKC2p}, for the case of the x-position reconstruction in the top row of pads. 
Similar results are obtained in the case of the bottom row, and for the determination of the y-position. 
The performance of the simple algorithm again yields a substantial gain with respect to the binary readout, improving from \SI{144}{\micro\metre} to a resolution of 20--40~\si{\micro\m}. 
Unlike the strip sensors the performance is not uniform across the surface, because the amplitude fraction as a function of the position flattens in the interior of the pads under the metal. 
The deterioration in the pad interior could likely be avoided by using a sensor with smaller metallized pads (and the same pitch). 

Similar performance is observed for the HPK B--2 pad sensor.


\subsubsection{Machine learning analysis}

As an alternative position reconstruction algorithm, a neural network (NN) based regression was trained to reconstruct the particle hit location within the sensor.
Similar NN specifically trained for AC-LGADs have been explored before~\cite{Siviero_2021}.
As a general motivation, there are situations in which the NN could outperform the reconstruction method described previously in terms of latency due to the fact that it is based on linear operations that are faster to evaluate on certain FPGAs and ASIC chips.
The functional form used for the amplitude ratio fits could become non-linear which is not ideal for front-end electronics. 

The NN was trained on data collected with the BNL 2020 sensor and used the position from the FNAL tracking telescope as a target for the reconstructed position. 
The selection of the data is identical to the selection used for the analysis described previously.
The NN was trained using a Keras-based~\cite{chollet2015keras} code framework.
The data were randomly split into three categories: training, testing, and validation. 
The training, testing, and validation categories are comprised of 80\%, 10\%, and 10\% of the total data, respectively.

The input variables for the NN are the measured amplitudes and time from each channel in the sensor.
The NN model is a fully connected dense NN (DNN) with three hidden layers.
The input layer is passed to a lambda layer that transforms each of the inputs such that their respective distributions have a mean of zero and standard deviation of one.
The number of nodes for each hidden layer was set to one hundred and the activation function for each of the hidden layers was set to the commonly used rectified linear unit (ReLU) function.
The output layer has only one value for its output and does not use an activation function.
The model is then trained using the Adam optimizer \cite{kingma2017adam} and by using the mean squared error as the loss function. 
The model is supervised to predict the position measured by the FNAL tracking telescope.
All hyper--parameters are scanned to determine the optimal values for each one.


The final results are compared to the telescope position to measure the position resolution from the NN method.
The performance is almost identical compared to the results obtained with the simplified method shown above.
Because the NN is trained to target the tracking telescope measurements, the position resolution will be limited by the resolution of the telescope. 
Therefore, the overall performance difference between the two reconstruction approaches can not be discerned at this stage.


\subsection{Time reconstruction}

Similar to position reconstruction, signal sharing can also be utilized to improve the time measurement of the AC-LGAD sensor.
The simplest approach for reconstructing the time for a given particle hit is to use the timestamp measured in the channel with the largest signal amplitude, using a constant fraction discriminator (CFD) algorithm at 20\% level to the leading edge of the signal.
However, this approach can lead to non-uniform time resolution across the sensor as the single-channel signal amplitudes are smaller at the boundary between two channels.

An improved method calculates timestamps according to the amplitude squared weighted average of multiple channels as follows
\begin{equation}
    t_{\rm{reco}} = \frac{\sum_{i}{a_i^2t_i}}{\sum_{i}{a_i^2}},
\label{eq:recotime}
\end{equation}
where $a_{i}$ and $t_{i}$ are the amplitude and time measurements of strip $i$.
This time reconstruction method can improve the performance at the interchannel boundaries.

As in DC-LGADs, the time resolution is in general dominated by two contributions: the Landau fluctuations in the ionization profile, which limits the resolution to at best 25-30 \si{\pico \second} for \SI{50}{\micro \m} sensors; and the jitter, which becomes subdominant at sufficiently high signal-to-noise ratio values. 
Since the jitter is independent in each channel, the multi-channel combination can reduce this contribution. However, since the ionization fluctuations are correlated across all channels, the Landau contribution is not affected by the combination.
In general, for large timing systems, variation in signal propagation time from signals originating at different positions can contribute to the time resolution. For the devices considered in this analysis, the variation in signal propagation time is small and does not significantly impact the time resolution. For larger devices, it may be necessary to add additional position-dependent delay corrections to Eq.~\ref{eq:recotime}.

\begin{figure}[htp]
\centering
\includegraphics[width=0.49\textwidth]{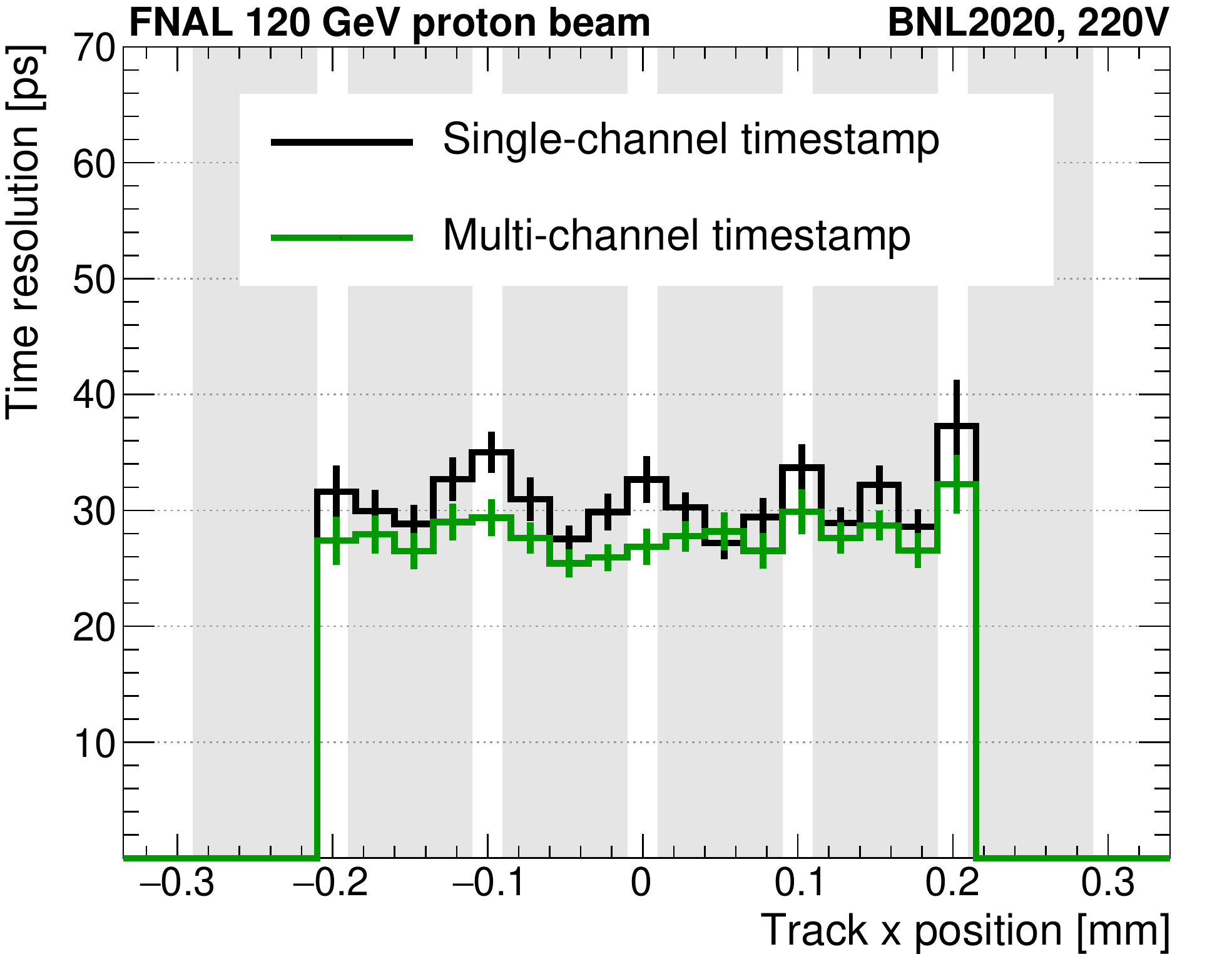}
\includegraphics[width=0.49\textwidth]{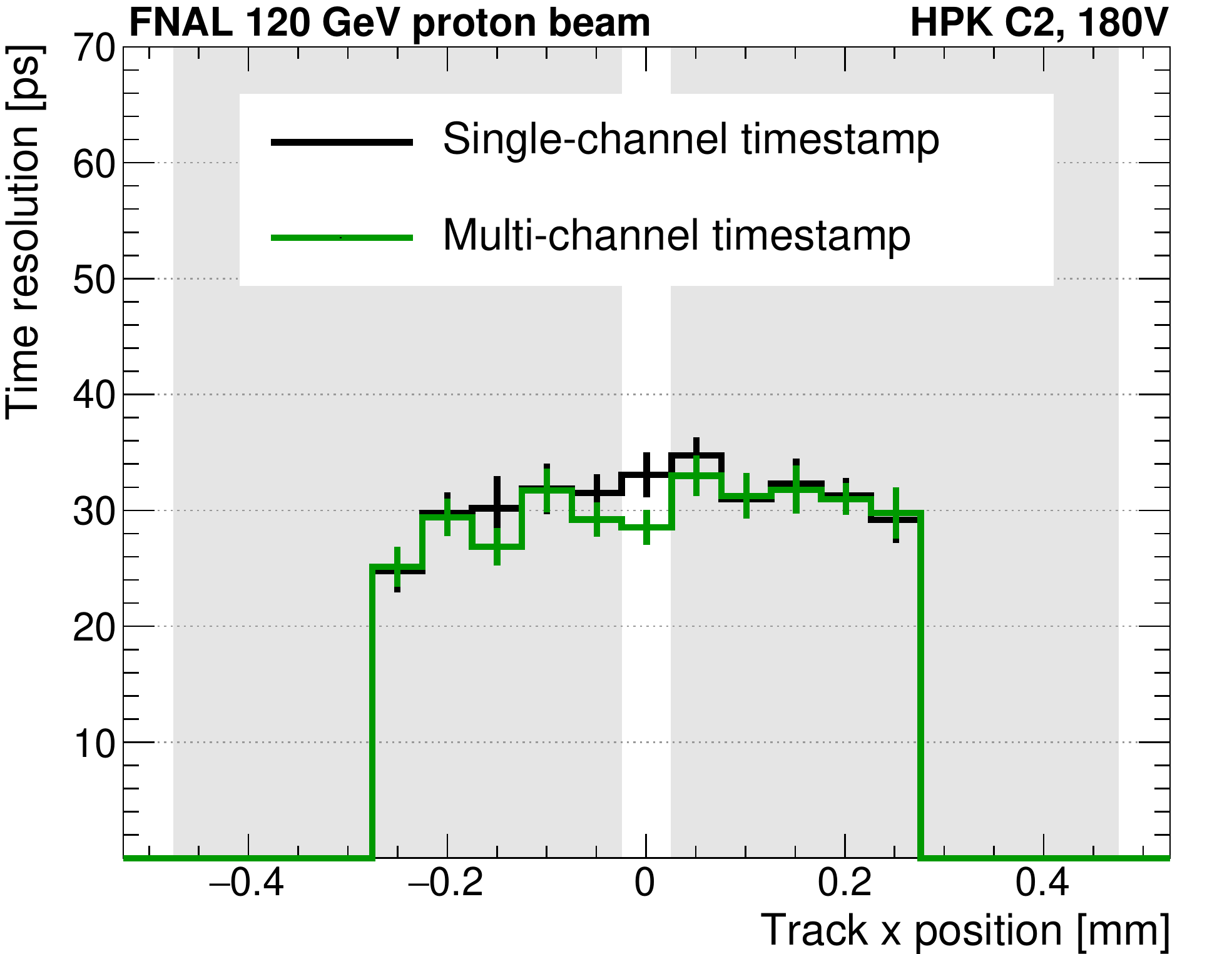}
\caption{Time resolution as a function of proton track hit position with different methods, for the BNL 2020 sensor (left) and HPK C--2 pad (right). The grey area indicates the metallized regions on the sensor surface. Errors represent the statistical uncertainty on the fits.
\label{fig:timeResvsX}}
\end{figure}

We show measurements of the time resolution for the BNL 2020 and HPK C--2 sensors as a function of the proton hit position along the axis perpendicular to the strip orientation in Fig.~\ref{fig:timeResvsX} using both the single and multi-channel timestamps.
The time resolution is measured by comparing the AC-LGAD timestamps with those from the Photek MCP-PMT. 
The time resolution of the Photek MCP-PMT has been measured to be about 10~ps, and this value is subtracted in quadrature from the time resolution measurements shown. 
For the strip sensors, when protons hits near the inter-strip boundaries, the signals are split evenly between the two adjacent strips. 
Since neither strip has a large signal in this case, the single-channel timestamps yield slightly worse performance. 
However, combining both signals in the multi-channel timestamp recovers the performance in the boundary regions, and ultimately delivers uniform resolution across the entire sensor at the \SI{30}{\pico\second} intrinsic floor for the strip sensors.
For the HPK C--2 sensor, the difference between the single and multi-channel timestamps performance is not as significant.


Similar results are obtained from all BNL strip sensors and both HPK pad sensors studied, and the time resolutions at optimal bias voltage reach the \SI{30}{\ps} Landau floor in all 6 sensors.


\begin{figure}[htp]
\centering
\includegraphics[width=0.49\textwidth]{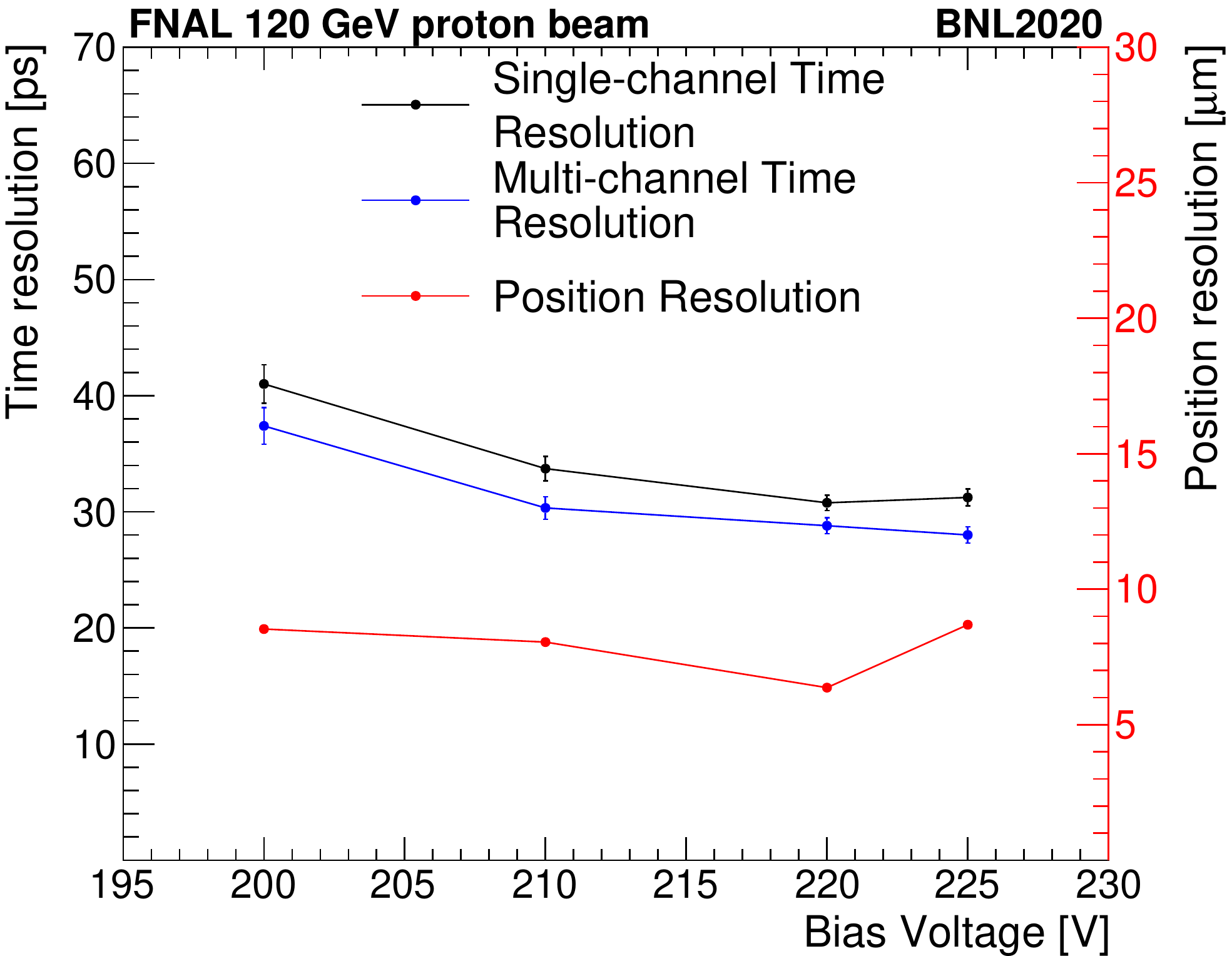}
\hspace{0.1cm}
\includegraphics[width=0.49\textwidth]{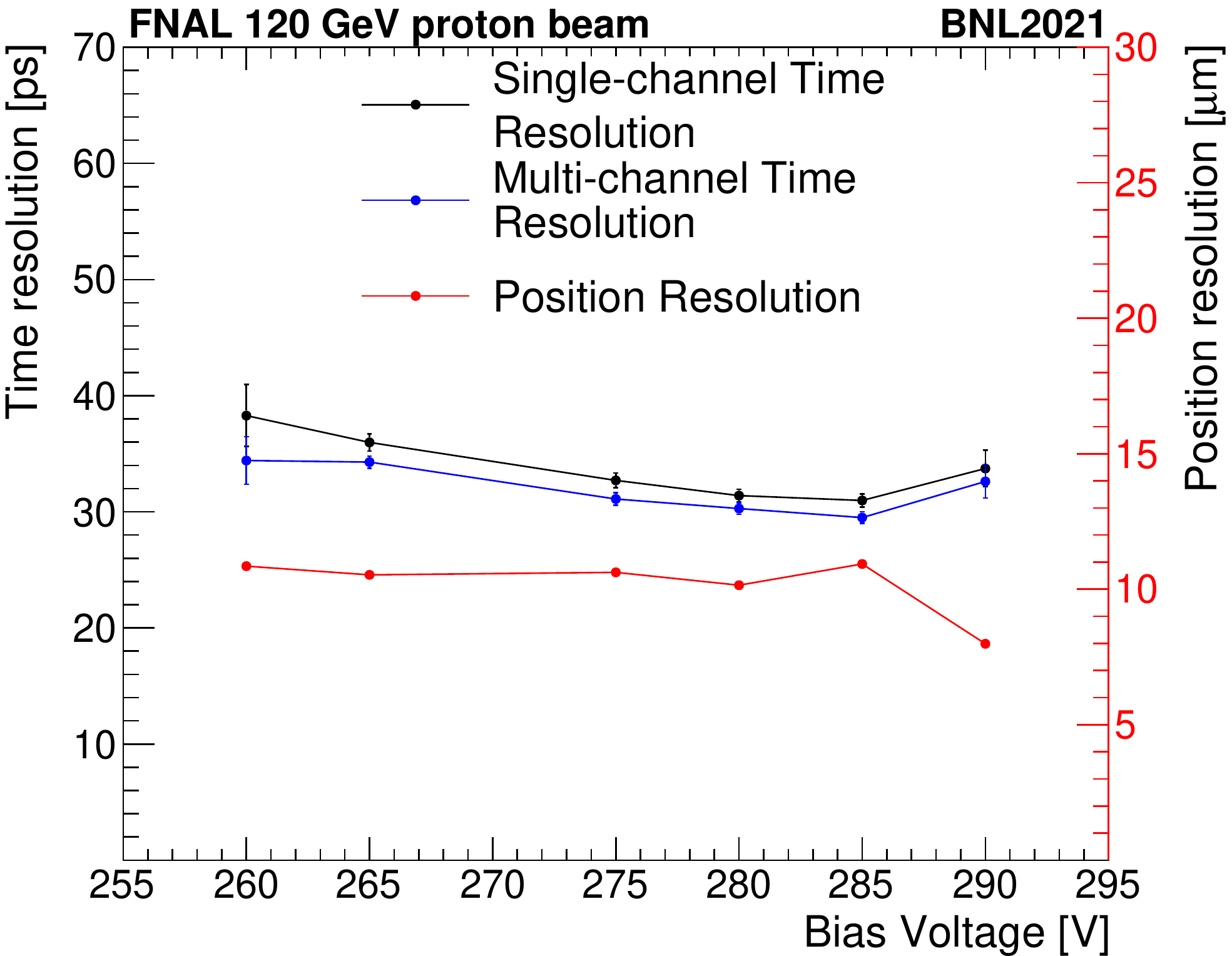}
\caption{Measurements of time and position resolution as a function of bias voltage for the BNL 2020 (left) and BNL 2021 medium pitch (right) sensor. Errors represent the statistical uncertainty on the fits.
\label{fig:biasscanBNL}}
\end{figure}

In Fig.~\ref{fig:biasscanBNL}, we show the time and position resolution, averaged across the BNL 2020 and BNL 2021 sensors, as a function of the bias voltage. The time resolution improves from about \SI{40}{\ps} to \SI{30}{\ps} as the bias voltage is increased. 

\subsection{Discussion}

In general, we find excellent overall performance from the sensors considered in this test beam campaign, as summarized in Table~\ref{table:SignalProp}. All four of the BNL AC-LGAD strip variants delivered position resolutions below 6--10 \si{\micro\m}, at the limit of the measurement sensitivity. Remarkably, the position resolution is at least 5--15 times finer than the resolution that would be obtained from binary readout alone, and is achieved by interpolating the signals between just the two leading strips in each event. The differences in the observed limits on the position resolution for the four BNL strip variants are not considered to be significant; rather, they likely arise due to differences in the telescope performance and alignment quality in each dataset.

All of the BNL strip sensors also reach time resolutions at the \SI{30}{\pico\second} limit that is intrinsic to \SI{50}{\micro\m} as in standard, i.e. DC-coupled, LGADs. 
Most of the timing performance is achieved by using only the timestamp from the leading amplitude channel, though the combination with the subleading channel ensures uniform resolution even in the interchannel boundaries where the signal in the leading channel is reduced. 
These sensors exhibit 100\% detection efficiency, even at the interchannel boundaries. 
The strips exhibit signals with most probable charge in the range of 10-30 \si{\femto \coulomb} in the primary strip, and 2-10 \si{\femto \coulomb} in the neighboring strips.

The BNL strips achieved good performance thanks to a resistive surface layer with sufficient resistivity to limit the signal sharing distance to 100-200 \si{\micro\m}, comparable to the pitch, so that the signals were concentrated in just two or three channels with high signal-to-noise ratios. 

The HPK sensors, on the other hand, exhibited signal sharing on larger distance scales. The longer signal sharing scale was well-suited for the geometry of the HPK \SI{500}{\micro\m} pad sensors. The pad sensors also delivered time resolution at the  \SI{30}{\pico\second} limit, and position resolutions approximately 7 times better than in binary readout. Comparing HPK resistivity variants C--2 and B--2, it is clear the signal sharing distance could be tuned to yield better performance also for  finer pitch sensors. 

The position resolution of the pad detectors was observed to be non-uniform, due to degradation for particles passing near the center of the pads. This can be understood from the signal sharing profile, which flattens for protons that pass through the inner regions of the metal pads. The signals in these events are absorbed almost fully by the primary pad, i.e. the pad that is directly hit by the proton. We believe that this effect could be mitigated by reducing the size of the metal electrodes from \SI{500}{\micro\m} to 200-300 \si{\micro\m}, which would allow more of the signal to reach the adjacent pads and recover uniform position resolution.

As a feasibility study, we also demonstrated a neural network reconstruction that was able to yield the same performance as the amplitude ratio method. Since both methods reach the limit of the telescope precision, we cannot discern an improvement in the performance from the neural network method.

\begin{table}[htp]
  \centering
  \caption{Performance summary for BNL strip and HPK pad detectors. A 10\% uncertainty is applied to the MPV signal amplitudes, representing the uncertainty in amplifier calibration. The position resolutions quoted for the strips are only upper limits since the measurements are limited by the resolution of the tracker reference. The HPK position resolution and all time resolution uncertainties represent the statistical error, only.}
  \begin{tabular}{ l | c | c | c | c }
  Name            & Pitch & Primary signal amp. & Position res. & Time res. \\
  Unit            & \si{\um} & \si{\mV}         & \si{\um}      & \si{\ps}  \\
  \hline\hline
  BNL 2020          & 100   & 101 $\pm$ 10  & $\leq$6    & 29 $\pm$ 1    \\ \hline
  BNL 2021 Narrow   & 100   & 104 $\pm$ 10  & $\leq$9    & 32 $\pm$ 1    \\
  BNL 2021 Medium   & 150   & 136 $\pm$ 13  & $\leq$11   & 30 $\pm$ 1    \\
  BNL 2021 Wide     & 200   & 144 $\pm$ 14  & $\leq$9    & 33 $\pm$ 1    \\ \hline
  HPK C--2          & 500   & 128 $\pm$ 12  & 22 $\pm$ 1 & 30 $\pm$ 1    \\ \hline
  HPK B--2          & 500   & 95  $\pm$ 10  & 24 $\pm$ 1 & 27 $\pm$ 1    \\ 
  \end{tabular}
  \label{table:SignalProp}
\end{table}



\section{Conclusions and Outlook}\label{sec:conclusions}

We present detailed studies of several AC-LGAD sensors exposed to 120 GeV proton beam at the Fermilab test beam facility. 
Several enhancements to the experimental setup were incorporated, such as improved DAQ that allows high speed readout of up to 7 channels, and significantly better position resolution measurement. 
The new readout allows multi-strip measurements and a combination of information that takes full advantage of signal sharing among neighboring channels.
A proton hit efficiency close to 100\% has been measured for individual strips as well as for a combination of adjacent strips. We demonstrate that uniform position resolution of around \SI{6}{\um} can be achieved with strip detectors. 
Uniform time resolution of around \SI{30}{\ps} can be achieved across the full surface of sensors with correctly optimized geometry.

These results are the first demonstration of simultaneous \SI{6}{\um} and \SI{30}{\ps} resolutions in a single sensor.
We demonstrate that the signal sharing observed in AC-LGADs can be utilized for much improved position reconstruction compared to standard silicon detectors, with the additional benefit of precision timing measurement. 

\acknowledgments
We thank the Fermilab accelerator and FTBF personnel for the excellent performance of the accelerator and support of the test beam facility, in particular M.~Kiburg, E.~Niner and E.~Schmidt. 
We also thank the SiDet department for preparing the readout boards by mounting and wirebonding the AC-LGAD sensors. 
Finally, we thank L.~Uplegger for developing the telescope tracker and a large part of the DAQ system.

This document was prepared using the resources of the Fermi National Accelerator Laboratory (Fermilab), a U.S. Department of Energy, Office of Science, HEP User Facility. 
Fermilab is managed by Fermi Research Alliance, LLC (FRA), acting under Contract No. DE-AC02-07CH11359.
This research is partially funded by the U.S.-Japan Science and Technology Cooperation Program in High Energy Physics, through Department of Energy under FWP 20-32 in the USA, and via High Energy Accelerator Research Organization (KEK) in Japan.
This work was also supported by the U.S. Department of Energy under grant DE-SC0012704; 
used resources of the Center for Functional Nanomaterials, which is a U.S. DOE Office of Science Facility, at Brookhaven National Laboratory under Contract No. DE-SC0012704;
supported by the Chilean ANID PIA/APOYO AFB180002 and ANID - Millennium Science Initiative Program - ICN2019\_044.
This research was partially supported by Grant-in-Aid for scientific research on advanced basic research (Grant No. 19H05193, 19H04393, 21H0073 and 21H01099) from the Ministry of Education, Culture, Sports, Science and Technology, of Japan. 


\bibliographystyle{report}
\bibliography{biblio}{}
\end{document}